\documentclass[fleqn,usenatbib]{mnras}

\usepackage{newtxtext,newtxmath}
\usepackage[T1]{fontenc}

\DeclareRobustCommand{\VAN}[3]{#2}
\let\VANthebibliography\thebibliography
\def\thebibliography{\DeclareRobustCommand{\VAN}[3]{##3}\VANthebibliography}

\usepackage{graphicx}	% Including figure files
\usepackage{amsmath}	% Advanced maths commands
\usepackage{graphicx}
\usepackage{xcolor}
\usepackage[export]{adjustbox}
\usepackage{pdflscape}
\usepackage{txfonts}
\usepackage[pdfpagelabels=false]{hyperref} % To add links in your PDF file, use the package "hyperref" with options according to your LaTeX or PDFLaTeX drivers.
\hypersetup{colorlinks=true,linkcolor=blue,citecolor=blue,filecolor=blue,urlcolor=blue,}
\usepackage{caption}
\usepackage{systeme}    % Systems of equations
\usepackage{subcaption} % use subfigure option
\usepackage{breqn}      % automatic breaking of long equations
\usepackage{rotating}
\usepackage{adjustbox}
\usepackage{tabularx}
\usepackage[flushleft]{threeparttable}
\usepackage{lipsum}
\usepackage{float}
\usepackage{placeins}

\newcommand{\revision}[1]{#1}

\title[The symphony of massive star variability]{The symphony of pulsations and binarity among massive stars using HERMES spectroscopy and TESS photometry}

\author[K. Thomson-Paressant et al.]{K. Thomson-Paressant$^{1}$\thanks{E-mail: Keegan.Thomson-Paressant@newcastle.ac.uk}, 
   D. M. Bowman$^{1,2}$,
   F. Nardini$^{1}$, 
   L. J. A. Scott$^{1}$,
   J. Bodensteiner$^{3}$,
   T. Shenar$^{4}$,
   \newauthor
   L. Mahy$^{5}$,
   G. Handler$^{6}$,
   N. Shitrit$^{4}$,
   I. Arcavi$^{4}$,
   M. Abdul-Masih$^{7,8}$,
   S. Simon-Diaz$^{7,8}$,
   P. Van Daele$^{1}$,
   \newauthor
   A. J. Kalita$^{1}$,
   L. Dennis$^{1}$,
   J. Henneco$^{1}$,
   A. Tkachenko$^{2}$,
   H. Sana$^{2,9}$,
   H. Van Winckel$^{2}$
\\
$^{1}$ School of Mathematics, Statistics and Physics, Newcastle University, Newcastle upon Tyne, NE1 7RU, United Kingdom\\
$^{2}$ Institute of Astronomy, KU Leuven, Celestijnenlaan 200D, 3001 Leuven, Belgium\\
$^{3}$ Anton Pannekoek Institute for Astronomy, University of Amsterdam, Science Park 904, 1098 XH Amsterdam, the Netherlands \\
$^{4}$ School of Physics and Astronomy, Tel Aviv University, Tel Aviv 6997801, Israel \\
$^{5}$ Royal Observatory of Belgium, Avenue Circulaire/Ringlaan 3, B-1180 Brussels, Belgium \\
$^{6}$ Nicolaus Copernicus Astronomical Center, Polish Academy of Sciences, ul. Bartycka 18, PL-00-716 Warszawa, Poland \\
$^{7}$ Instituto de Astrofısica de Canarias, C. Vıa Lactea, s/n, 38205 La Laguna, Santa Cruz de Tenerife, Spain\\
$^{8}$ Universidad de La Laguna, Dpto. Astrofısica, Av. Astrofsico Francisco Sanchez, 38206 La Laguna, Santa Cruz de Tenerife, Spain \\
$^{9}$ Leuven Gravity Institute, KU Leuven, Celestijnenlaan 200D, 3001 Leuven, Belgium\\
}

\date{Accepted 2026 July 31. Received 2026 July 28; in original form 2026 June 8.}

\pubyear{\the\year{}}

\begin{document}
\label{firstpage}
\pagerange{\pageref{firstpage}--\pageref{lastpage}}
\maketitle

\begin{abstract}
A wide range of variability mechanisms exist among intermediate mass and massive stars, which are not yet fully understood. Using complementary data sources for a large population of B- and O-type stars, we aim to study the prevalence and interplay of different types of variability, including binarity, pulsation, and rotation, to prepare for future modelling. To this end, we analyse high-resolution HERMES spectra and 2-min cadence TESS photometry and characterise the diverse variability observed within a population of 873 O- and B-type stars. The spectroscopic data were normalised using machine-learning techniques, compared to a grid of synthetic TLUSTY spectra to determine stellar parameters, and used to identify radial velocity variability. Photometric time series were analysed using standard frequency analysis methods to detect pulsations and rotational modulation signatures. We find that more than 93~per~cent of the sample exhibits photometric variability. Photometric variability caused by pulsations is identified in 82~per~cent of the sample, with dominant contributions from $\beta$~Cep and slowly pulsating B-type stars, as well as stochastic low-frequency variability. Based on a limited number of spectroscopic epochs, at least 14~per~cent of the stars show evidence of binarity, including both eclipsing and spectroscopic systems. This work represents one of the largest homogeneous surveys of variability for intermediate-mass and massive stars in the Northern hemisphere, and complementing similar efforts in the Southern hemisphere. It provides a statistical framework for future studies of stellar structure and evolution, particularly in the context of asteroseismology.
\end{abstract}


%%%%%%%%%%%%%%%%%%%%%%%%%%%%%%%%%%%%%%%%%%%%%%%%%%

%%%%%%%%%%%%%%%%% BODY OF PAPER %%%%%%%%%%%%%%%%%%

\section{Introduction}
Intermediate- and high-mass stars play a fundamental role in stellar and galactic evolution, spanning regimes that contribute both to local stellar physics and to the large-scale evolution of galaxies. Through their intense radiation fields, strong stellar winds, and dramatic deaths as supernovae or gamma-ray bursts they greatly impact their environment \citep{bromm2004}. The heavier elements they produce in their cores, which are then distributed via mass loss or via explosive events at the ends of their lives, are a major contributor to chemical enrichment, which can both inhibit or promote the formation of new generations of stars \citep{massey2003,langer2012}. Finally, as the progenitors of both black holes and neutron stars, whose merger events produce gravitational waves, massive stars shed new light on the structure and expansion of the cosmos, and allow for the testing of novel physical prescriptions for gravity \citep[e.g.][]{abbott2017b,abbott2017,laplace2020}. While the most massive stars dominate feedback through winds and explosive endpoints, intermediate-mass stars provide an important link between low- and high-mass stellar structure regimes. In particular, stars in the late-B and A spectral range exhibit a transition in internal structure and variability properties, making them valuable laboratories for studying transport processes, pulsation driving, rotation, and magnetism across the upper main sequence \citep{aerts2021,bowman2020,kurtz2022}.

Intermediate- and high-mass stars are governed by a complex interplay of several physical processes such as convection, rotation, mixing, magnetic fields, and mass loss, many of which remain imperfectly understood. As such, acquiring a greater knowledge of each of these processes is essential, not only for achieving a better picture of the life cycles of intermediate- and high-mass stars, but also for informing our broader understanding of stellar structure and evolution across the Hertzsprung--Russell (HR) diagram \citep{meynet2000,aerts2021,kurtz2022}.

Binarity plays a key role in the evolution of stars (e.g. \citealt{sana2012,demink2013}), altering their internal structure, angular momentum evolution, evolutionary pathways, and end products through processes such as tidal interactions, mass transfer, and mergers \citep[see][for reviews]{langer2012,marchant2024}. Pulsating eclipsing binaries in particular represent some of the best laboratories for improving stellar evolution theory \citep[see][for a review]{southworth2025}. Observational studies of O- and B-star populations have demonstrated that multiplicity is extremely common, with inferred binary fractions ranging from approximately 25 to 80~per~cent in O- and B-type stars depending on stellar mass, orbital separation sensitivity, and the properties of the parent population (e.g. \citealt{moe2017, offner2023}). 
%Recent dedicated surveys have further shown that the binary properties depend on both environment and metallicity. 
For example, studies of early B-type stars in galactic open clusters reveal high multiplicity fractions ($\geq 50$~per~cent) and a significant population of close binaries \citep{banyard2022, frost2025, nardini2025}. Surveys at subsolar metallicity, such as in the LMC, report similarly elevated binary fractions ($\geq 60$~per~cent) among massive stars \citep{sana2012,sana2013,sana2025,villasenor2025,dunstall2015}. These results indicate that binarity is not only prevalent but also a fundamental characteristic of massive-star populations across a wide range of environments. Consequently, accurately identifying and characterising binary systems is essential for interpreting observed stellar properties, disentangling binarity-induced variability from intrinsic stellar phenomena, and placing meaningful constraints on stellar evolution models (see \citealt{southworth2025}).

One of the most informative methods for probing the physics of stars is through the study of their variability. Variations in brightness and spectral lines arise from a variety of mechanisms, including binarity, rotational modulation, pulsation, mass loss, and magnetic activity. Each of these processes offers insight into different layers or aspects of a star's structure. This is particularly true across the \revision{O- and B-star regimes}, where pressure modes, gravity modes, and rotational variability coexist and probe different regions of stellar interiors, enabling comparisons across stellar mass and evolutionary state \citep{aerts2010,bowman2020}. Traditional observational methods, such as photometry and high-resolution spectroscopy, have long enabled the detection and characterisation of these phenomena. More recently, the study of stellar oscillations via asteroseismology (see \citealt{aerts2010}) has emerged as a powerful tool to investigate internal stellar structure and infer stellar parameters with unprecedented precision --- see review by \citet{kurtz2022}. Thanks to recent space-based missions such as the Convection, Rotation and planetary Transits \citep[CoRoT;][]{auvergne2009} satellite, \emph{Kepler} \citep{borucki2010}, and the Transiting Exoplanet Survey Satellite \citep[TESS;][]{ricker2015}, our ability to utilise this technique has been greatly expanded (see reviews by \citealt{chaplin2013, hekker2017, bowman2020, aerts2021}).

In this study, we have performed a combined photometric and spectroscopic analysis of 873 O- and B-type stars in the Northern hemisphere, using recent TESS space mission light curves and multi-epoch high-resolution spectra from HERMES at the Mercator telescope (Section~\ref{sec:data-analysis}). 
%In this study, we have performed a combined photometric, using recent TESS space mission light curves, and multi-epoch high-resolution spectroscopic analysis of 873 OB-type stars in the Northern hemisphere (Section~\ref{sec:data-analysis}). 
We present here the demographics of variability caused by pulsations and binarity displayed in the sample determined using both photometry and spectroscopy (Section~\ref{sec:distribution}). A few studies of smaller samples have been performed previously for massive stars in the Southern hemisphere \citep[e.g.][]{burssens2020}, and so in this work we complement this with a (much larger) sample of OB-type stars in the Northern hemisphere. The goal of this work is to identify stars with variability caused by pulsations (Section~\ref{sec:pulsations}), binarity (Section~\ref{sec:radvel}), as well as candidate magnetic stars based on rotational modulation (Section~\ref{sec:magnetism}), 
% as well as mass loss and rapid rotation (Section~\ref{sec:oe-be}), 
which in future work can be studied using binary modelling \citep[e.g.][]{zapartas2026}, in-depth magnetic characterisation \citep[e.g.][]{erba2024}, and forward asteroseismic modelling (e.g. \citealt{burssens2023, vanlaer2025, fritzewski2025}). 
%Pulsating eclipsing binaries are by far the most valuable targets since the model-independent masses and radii from binary modelling are powerful constraints on subsequence asteroseismic inference --- see \citet{southworth2025} for a review.

\section{Data analysis}
\label{sec:data-analysis}

\subsection{Target selection}
\label{subsection:Target-Selection}

At the inception of this project, we used SIMBAD \citep{wenger2000} to construct a sample of 4121~stars with a spectral type of O or B and bright enough (i.e. $V \leq 14$~mag) to prepare a series of TESS Guest Investigator (GI\footnote{\url{https://heasarc.gsfc.nasa.gov/docs/tess/approved-programs.html}}) proposals throughout cycles 3-8. These targets are typically located in star-forming regions in the Milky Way. However, the sampling of the sky by the TESS mission is not regular nor uniform. In cycles 1 and 2 (2018-07-25 to 2020-07-04), TESS observed 13 sectors (each lasting 28~d) in each ecliptic hemisphere using four cameras with a combined field of view of $24\times96^\circ$. In cycles 3 and 4 (2020-07-05 to 2022-09-01), TESS observed 13 sectors in the South, 11 in the North, and a further 5 sectors covering the ecliptic plane itself. In cycles 5 and 6 (2022-09-01 to 2024-10-01) TESS covered 9 southern sectors, 16 Northern sectors, and 3 ecliptic sectors. Cycle 7 (2024-10-01 to 2025-09-15) covered 3 Northern, 2 ecliptic and 8 southern sectors, and cycle 8 is currently ongoing and due to end on 2026-09-07.

Due to the evolution of observing strategy over the course of the TESS mission, the cadence and duration of the assembled light curves is star dependent. For example, some stars have continuous light curves spanning up to 1~yr because they lie in one of the continuous viewing zones (see \citealt{ricker2015}), whereas the majority of stars only have 1~sector of TESS data in each cycle (approximately 27~d). From the large sample of 4121 OB-type stars with $V < 14$~mag, a total of 2248 were selected to be observed with a short cadence of 2~min in the context of several successful GI proposals\footnote{proposal IDs: G03059, G04074, G05036, G06037, and G07037 (PI: Bowman).}.

\begin{figure}
    \centering
    \includegraphics[width=0.98\linewidth]{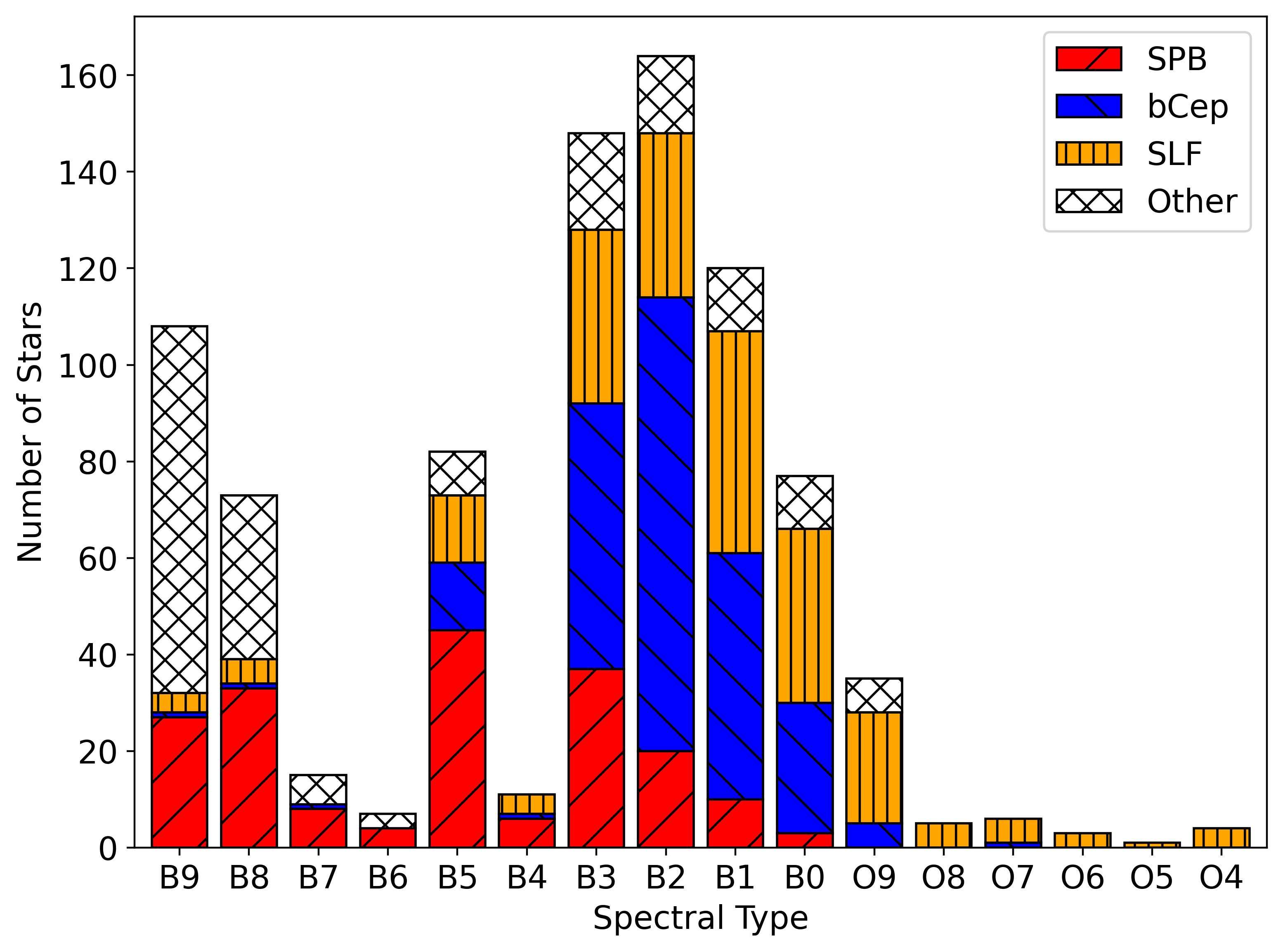}
    \caption{Distribution of spectral types from SIMBAD for the stars in our sample, classified by pulsator class from our analysis of TESS photometry, described in detail in Section~\ref{sec:tess}.}
    \label{fig:spt}
\end{figure}

% \begin{figure}
%     \centering
%     \includegraphics[width=0.98\linewidth]{Figs/CMD.png}
%     \caption{\revision{Colour-magnitude diagram using values retrieved from \emph{Gaia}, with the stars in the sample coloured with respect to their SIMBAD spectral type. A representative set of stars have been added to the background in grey.}}
%     \label{fig:cmd}
% \end{figure}

\subsection{HERMES spectroscopic data}
\label{sec:spectroscopy}

To complement the TESS observations, we initiated a large spectroscopic follow-up programme with the 1.2-m Mercator telescope\footnote{\url{https://www.mercator.iac.es/instruments/}}, which is located on La Palma, Spain. The latitude and technical specifications of the observatory imposed a brightness requirement of $V < 12$~mag and restrictions to targets mostly in the Northern hemisphere (i.e. $\delta > -30^{\circ}$). We also removed any stars from the follow-up programme that already had previous observations in the Mercator archive. The large programme with the Mercator telescope to assemble these spectra ran between 2022-2025 (programme ID~123; PI: Bowman). The final sample presented and analysed in this paper with TESS data and new Mercator spectra is 873 stars.

The spectroscopic data in this work were assembled with the High Efficiency and Resolution Mercator Echelle Spectrograph (HERMES; \citealt{raskin2011,raskin2014}). The HERMES instrument is a fibre-fed prism-cross-dispersed echelle spectrograph based on a white-pupil design, a CCD with 2048$\times$4608 pixels, and covers the 3800-9000 \AA{} wavelength range with a resolving power of $R = 85000$ \citep{raskin2011,raskin2014}. The spectra were reduced by the HERMES data reduction software (HERMES-DRS, v.7.0), which includes bias and dark correction, flat-fielding, wavelength calibration, cosmic ray removal, and correction for barycentric motion. 

The goal of our spectroscopic programme was to acquire three epochs per star, two separated by 1-2 months and a third occurring about a year later. Thus, both short- and long-period radial velocity \revision{(RV)} variability should be detectable. Of the 873 stars in our sample, 138 successfully fulfilled these criteria (with a further 32 that have at least 3 spectra but do not fulfil the scheduling criteria), 464 targets have two out of three epochs requested, and 409 targets have only a single epoch to date. Nonetheless, we retain the entire sample of 873 stars for spectroscopic analysis. 
%Moreover, 32 stars have at least 3 epochs but do not successfully fulfil the criteria. Finally, for the 294 targets that have 2 spectroscopic epochs available, the mean separation between observation epochs is 222 days (min: 1, max: 1122, median: 76 days).
\revision{Signal-to-noise ratios (S/N) for the various spectra utilised in our sample range from 20 to 385, with a mean value of 105, calculated at the 65th spectral order. The distribution of spectral types contained within the sample are shown in Fig.~\ref{fig:spt}.}
%, with the same sample also represented in the colour-magnitude digram displayed in Fig.~\ref{fig:cmd} using values from \emph{Gaia}.}

\subsubsection{Visual inspection to identify SB2s and emission line stars}

The reduced HERMES spectra were assessed to look for a variety of features, including spectral line variability, chemical peculiarity, binarity, and emission features. We first visually inspected the regions corresponding to H$\alpha$ and He\,{\sc i} lines at 6563 and 6678~\AA{}, respectively, to look for emission features typical of either OBe stars or blue supergiants (BSG). We simultaneously sought to identify double-lined spectroscopic binary (SB2) systems as, together with the emission-line stars, these targets would need to be treated differently in the following steps. For completeness, we also compared our sample with the `SBX' database of spectroscopic binaries \citep{pourbaix2004,merle2026}, and took note of any overlap. This database does not explicitly classify targets as either SB1 or SB2, but we took note of the 38 targets that were included in SBX for the future analysis steps. 

Inspection of the data resulted in the identification of 22 SB2 systems, 131 OBe stars, and 33 BSGs (determined through a combination of P\,Cygni-like emission features in the H$\alpha$ line region and the spectral type provided from SIMBAD). This leaves us with 687 non-emission line and non-SB2 stars for subsequent analysis. We leave the analysis of emission-line stars and SB2 systems for future work.
%The results of these analyses are presented in Sec.~\ref{sec:radvel} for the SBs and Sec.~\ref{sec:oe-be} for the OBe stars. 

\subsubsection{Normalisation}

After having identified and separated emission-line stars and SB2 systems out from the sample, we normalised all the remaining spectra. Instead of performing normalisation manually on a star-by-star and spectrum-by-spectrum basis, the HERMES spectra were continuum normalised automatically utilising the {\sc SUPPNet}\footnote{\url{https://git.io/JqJhf}} software package \citep{suppnet}. {\sc SUPPNet} is a neural network designed to predict the location of the pseudo-continuum in stellar spectra, having been trained on both synthetic and real spectra of stars spanning the O to G spectral types, and with a wide range of variability included (e.g. pulsation, chemical peculiarity, and emission features). 
% However, the stars with emission lines required a different treatment. Specifically, some of the input parameters of {\sc SUPPNet} auto-normalisation needed to be adjusted to account for the different spectral line shapes. This involved informing the neural network to expect emission lines, by changing the `weights' option from `active' (default) to `emission', and slightly increasing the sampling width (from the default 0.05 to 0.3). 

\subsubsection{Cross correlation}

After normalisation, next we followed the procedure presented in \citet{shenar2019}, \citet{dsilva2020}, and \citet{nardini2025}, which was adapted from \citet{zucker2003}, and performed a cross-correlation\footnote{\url{https://github.com/TomerShenar/Cross-correlation}} technique on the continuum-normalised spectra for each star. Using a set of spectral lines as reference points, the spectra were co-added and weighted based on their respective \revision{S/N}. To this end, we utilised the He\,{\sc i} absorption lines located at 4472, 5876, and 6678 \AA{} as the reference points to anchor the cross-correlation across all the epochs available for a given star.
% while for the targets showing broad emission features we had to be more selective in order to avoid contamination. For these stars, we relied more often on metal lines such as the FeI lines at 6345, 6370, 6385, and 7512 \AA{}, as well as the OI lines at 7773 and 8446 \AA{}, however these stars proved difficult to work with and have been treated separately (see Sec.~\ref{sec:TvG} for more details).}
The benefits of this cross-correlation process are two-fold: (i) it can be used to determine the reference point for calculating RV variability; and (ii) it provides a combined high-S/N spectrum for stellar parameter determination. 
%As part of this process, the spectra are shifted in wavelength to account for any RV variability, using the spectrum with the highest S/N as the template. 

%\subsubsection{RV variability}

\revision{We determined RV measurements for each spectrum per target using the co-added spectrum as a template.}
%These RVs are calculated using cross correlation, based on work by \citet{zucker2003} and further developed by \citet{shenar2019} and \citet{dsilva2020}. 
With at least two spectroscopic epochs for a given star we can calculate a $\Delta \mbox{RV}$ (i.e. the maximum absolute variation in RV seen across available spectra). Additional epochs naturally lead to higher precision and better constraints on this value; with only two spectra available we can at least determine a lower limit. From the sample, we identify 277 of the 377 stars with two or more epochs (after removing targets with emission and SB2 systems) as being RV variables. This was assessed by determining whether $\Delta \mbox{RV} > \sigma\Delta \mbox{RV}$ (i.e. the maximum variation in RV was larger than its error). This RV variability could result from a variety of processes including pulsation and binarity, but disentangling these effects is difficult with only spectroscopy, especially when only a few epochs are available. This measurement is only utilised for classifying stars in our sample as RV variables, and is separate from the two criteria utilised for determining binarity, as discussed further down in this section.

Following previous spectroscopic studies of massive stars \citep{sana2012,bodensteiner2021,banyard2022,mahy2022,nardini2025}, we apply two RV variability criteria to separate single stars from likely binary systems. We emphasise that identifying binaries from RV variability only yields the observed binary fraction rather than the intrinsic binary fraction, which is obtained by correcting for observational and astrophysical biases \citep[e.g.][]{sana2012,sana2013}.

In our sample of 687 stars (after removing SB2 systems and emission-line stars), only 377 stars have two or more spectroscopic epochs, and only 134 stars have three or more spectroscopic epochs. Therefore, owing to the limited number of epochs, it is not possible to determine orbital periods from the spectra alone. Moreover, for the majority of targets, only a single available epoch means it is impossible to determine any RV variability. As a mentioned previously (c.f. Section~\ref{sec:spectroscopy}), the RVs for the 377 stars with at least two spectroscopic epochs are determined using their co-added spectra as templates, with weights based on the S/N of the spectra. 

The first criterion to identify a candidate binary system is based on whether the RV variability is statistically significant, which is defined as
\begin{equation}\label{eq:rv1}
    \frac{|v_i - v_j|}{\sqrt{\sigma_i^2+\sigma_j^2}} > 4.0,
\end{equation}
where $v_i$ and $v_j$ are two distinct RV measurements, and their corresponding uncertainties $\sigma_i$ and $\sigma_j$, respectively. We adopt an identical threshold for the right-hand side of Eq.~\ref{eq:rv1} as previous studies to limit the impact of false-positives \citep{sana2012,bodensteiner2021,banyard2022,mahy2022,nardini2025}. From the 377 stars with multi-epoch spectroscopy that show RV variability, there are 166 targets that fulfil this criterion.

The purpose of the second criterion is to remove systems with RV variability that may be caused by a physical mechanism other than binarity (e.g. pulsations). To achieve this it is generally assumed that a true binary system has a minimum variation in its RV time series, $\Delta \mbox{RV}$, that is larger than some threshold. The second criterion to identify a binary system is thus
\begin{equation}\label{eq:rv2}
    \Delta \mbox{RV} = \left| v_i - v_j \right| > C ~ ,
\end{equation}

\noindent where $C$ is typically taken as 20~km\,s$^{-1}$ \citep{sana2013,bodensteiner2021,banyard2022,mahy2022,nardini2025}. 

Previous studies have set this threshold at $C = 20$~km\,s$^{-1}$ for O-type stars in the Galaxy, \revision{as well as the Large and Small Magellanic Clouds \citep[LMC and SMC, respectively;][]{sana2012, sana2013, sana2025,simondiaz2024}} because such stars do not typically exhibit high-amplitude coherent pulsations. However, exceptions exist, with the massive single star $\zeta$~Oph having intrinsic RV variability caused by pulsations exceeding 20~km\,s$^{-1}$ \citep{kalita2025}. On the other hand, the choice of this threshold is less clear for B-type stars \citep[see e.g.][]{villasenor2025,bodensteiner2025, britavskiy2025}, especially those in the Milky Way because such stars are commonly pulsators. \citet{banyard2022} and \citet{nardini2025} use the same threshold of $C = 20$~km\,s$^{-1}$ for different samples of galactic B-type stars in clusters and briefly discuss whether the threshold should be different. This is because the typical RV variability of pulsations in galactic B-type stars can commonly exceed 20~km\,s$^{-1}$ value (see e.g. \citealt{aerts2009,simondiaz2024}). Moreover, with only a limited set of spectroscopic epochs it is difficult to distinguish pulsations and binarity in galactic B-type stars. 

% \subsubsection{Chemical peculiarity}

% We also analysed the 5200 \AA{} region to identify the presence of a characteristic depression in the continuum together with the strength of the Si\,II lines at 5041 and 5055 \AA{}. These have been shown to act as reliable indicators for magnetism in chemically peculiar (CP) stars \citep{maitzen1976,kupka2003,kupka2004,khan2007,stigler2014,thomson2024}. This magnetic indicator feature has been found in stars with spectral types ranging from mid-F to mid-B, and thus have a reasonable overlap with our sample of stars. For the hottest stars in our sample (earlier than spectral type B4, which is the empirical upper limit for such a feature reported by \citealt{hummerich2020}), identifying CP stars becomes difficult. This is because most of the elements typically utilised to identify CP stars are ionised and difficult to detect due to the extreme atmospheric conditions strongly affecting atomic diffusion \citep[(i.e. mass loss, increased ionisation, macroscopic mixing processes, etc.][]{michaud1970,alecian2019}. For such hot stars, one would need to rely on differences in abundances of lighter elements (e.g. helium, carbon, nitrogen, oxygen, etc.) to classify CP stars, which is an aspect beyond the scope of this work. 

\subsubsection{Stellar parameter estimation}

\revision{Once again using} the co-added spectrum, and the RV shifts determined in the previous step, we also inferred the effective temperature ($T_{\text{eff}}$), surface gravity ($\log\,g$), and projected surface rotational velocity ($v\,\sin\,i$), of each star by comparing the co-added spectrum to a grid of synthetic spectra. The procedure follows previous work with similar goals and sample properties (see \citealt{bodensteiner2021}). Considering the range of spectral types in our sample, we elected to utilise the {\sc TLUSTY} grid of models \citep{hubeny1995,lanz2003,lanz2007}.
%shown to provide good constraints for stars with spectroscopic variability caused by binarity (e.g. \citealt{banyard2022, nardini2025}). 
The {\sc TLUSTY} models are separated into two grids, one for the O-type stars \citep{lanz2003} and one for the B-type stars \citep{lanz2007}, 
%both covering the UV (900-2000 \AA{}; not covered by HERMES) and 
covering the visible wavelength regime (3000-7500 \AA{}) utilised by HERMES. The parameter space of each grid in the visible wavelength regime is shown in Table~\ref{tab:tlusty}. For our sample of stars, we elected to fix the metallicity to solar ($Z/Z_\odot = 1$), with microturbulence set to $v_{\text{mic}} =$ 2 km\,s$^{-1}$ for the B-star grid and 10 km\,s$^{-1}$ for the O-star grid. Additional models were also calculated between 9000 and 15\,000~K, in steps of 1000~K, and with surface gravity $\log g$ ranging between 3.0 and 5.0 in steps of 0.25 dex, to complement the standard TLUSTY B-star grid and cover the lowest-temperature end of the sample. To account for line broadening due to rotation, the entire grid was convolved using the \textsc{ROTBROAD}\footnote{\url{https://pyastronomy.readthedocs.io/en/latest/pyaslDoc/aslDoc/rotBroad.html}} package (version 0.22.0), to generate additional models with $v~\sin~i$ values ranging from 0 to 500 km~s$^{-1}$ with steps of 20 km~s$^{-1}$. For calculating the stellar parameters, we relied on fitting the synthetic spectra to the Balmer line series, as well as He\,{\sc i} lines at 5875, 6678, and 7065 \AA{}, and He\,{\sc ii} line at 4685 \AA{}.
%for stars classified as B0 or earlier. 
The best-fitting stellar parameters, and their respective errors, were determined using a $\chi^2$ minimisation procedure, as described in \citet{Bodensteiner2023}. In short, this is done by comparing the co-added spectrum of a given star to each model in the grid and calculating
\begin{equation}
    \chi^2 = \frac{1}{S/N} \sum_{i=1}^{n}({\rm obs}_i-{\rm model}_i)^2,
\end{equation}
where ${\rm obs}_i$ and ${\rm model}_i$ are the observed and model spectra respectively at a given wavelength {i}, with $n$ being the total number of wavelength values tested. 

We compared the best-fitting parameters based on the spectra normalised using \textsc{SUPPNet} with those resulting from a manual continuum normalisation for a small subset of stars that span the full O4-B9 spectral range of the sample. We find that the parameter values determined from the two different sets of normalised spectra were within the confidence intervals of each other and are thus consistent. This demonstrates reliability in the automated method via \textsc{SUPPNet}, which proved extremely beneficial for performing this normalisation step on such an extensive sample. Several examples of the comparison between \textsc{SUPPNet} and manual normalisation are shown in Fig.~\ref{fig:contnorm} in the Appendix.

\begin{table}
\centering
\caption{Parameter space for the two TLUSTY model grids, including the additionally calculated models, with the grid step size indicated in parentheses for parameters we iterated over.}
\begin{tabular}{c c c}
\hline\hline
 & B-star & O-star \\
 \hline
$T_{\text{eff}}$ (K) & 9000 -- 30\,000 (1000) & 27\,500 -- 55\,000 (2500) \\
$\log g$ (dex) & 1.75 -- 4.75 (0.25) & 3.00 -- 4.75 (0.25) \\
$Z/Z_\odot$ & 0.01 -- 2 & 0 -- 2  \\
$v_{\text{mic}}$ (km\,s$^{-1}$) & [2;10] & 10 \\
$v\sin i$ (km\,s$^{-1}$) & 0 -- 500 (20) & 0 -- 500 (20) \\
\hline\hline
\end{tabular}
\label{tab:tlusty}
\end{table}

\subsection{Gaia data}

Primarily to verify the stellar parameters we acquired from comparison with the TLUSTY models, we also retrieved effective temperatures, surface gravities, and projected rotational velocities where possible from the \emph{Gaia} DR3 archive \citep{gaia2023}. We choose the values determined using the Extended Stellar Parametrizer for Hot Stars \citep[ESP-HS;][]{gaia2023c}. The ESP-HS includes corrections to the stellar parameters determined for stars with effective temperatures $T_{\text{eff}} \gtrsim 7500$~K, which is the case for our entire sample. Such parameters have been shown to be less reliable when determined using the standard procedures \citep[i.e. the General Stellar Parametrizer from Photometry, GSP-Phot;][]{fouesneau2023}. 
% (i.e. the General Stellar Parametrizer from Photometry, GSP-Phot;). 
The comparisons between the ESP-HS and TLUSTY methods of stellar parameter determinations are discussed in Section~\ref{sec:TvG}.

\subsection{TESS photometric data}
\label{sec:tess}

%Thanks to the success of recent space missions, there is an abundance of high-quality space telescope photometry which we can draw from to provide further insight into the features of the stars in the sample. 
In this work, we use time-series photometric data from the NASA TESS space mission \citep{ricker2015}, which are made publicly available on the Mikulski Archive for Space Telescopes (MAST) at the Space Telescope Science Institute\footnote{\url{https://archive.stsci.edu/missions-and-data/tess}}. 

For the vast majority of the targets in the sample ($>$96\%), 2-min cadence TESS data was available from the MAST database, which formed the basis upon which our photometric analysis was performed. When retrieving data from MAST, the light curves are provided with two different detrending methods, Single Aperture Photometry (SAP) and Pre-search Data Conditioning SAP (PDC-SAP). In almost all cases we relied on the PDC-SAP light curves, which are produced by the NASA SPOC pipeline \citep{jenkins2016}. Historically these detrended light curves have been shown to generally provide better results for measuring the photometric variability of massive stars (e.g. \citealt{burssens2020, bowman2022a, burssens2023}), except sometimes in the case of the brightest stars (e.g. \citealt{southworth2022}). In our sample, which only had 26 targets with $V < 5$ mag, the PDC-SAP remained the more reliable choice despite testing both methods.

For the portion of the sample for which the 2-min cadence data is not available (25 stars), we instead extracted light curves from the TESS full frame images (FFI), using the \textsc{Lightkurve} \citep{lightkurve} and \textsc{TESScut} \citep{brasseur2019} packages to select a 40$\times$40 pixel image around each target from which to extract the light curves. The latter were detrended using a principle component analysis (PCA) method, performed by excluding a 10$\times$10 pixel square centred on the target star and using the remaining pixels in the image as regressors (see e.g. \citealt{scott2026}). 

All the light curves, regardless of their origin, were then subjected to a Fourier analysis to identify significant frequencies associated with one or several variability mechanisms. We calculated frequency spectra using the modified generalised Lomb-Scargle (LS; \citealt{lomb1976, scargle1982}) periodogram from the \textsc{Astropy} \citep{astropy2013,astropy2018} package. This not only allowed us to identify the existence of significant peaks that correspond to pulsation mode frequencies, but also to infer the rotation and/or orbital frequencies of stars (see Section~\ref{sec:pulsations}). 

Generally, we use the same classification criteria as \citet{burssens2020}. For example, stochastic low-frequency (SLF) variability is defined as an excess of power in the low frequency domain not related to coherent modes \citep{bowman2019}, slowly pulsating B-type (SPB) stars are main-sequence B-type stars that pulsate in coherent non-radial g~modes with frequencies between about 0.2 and 4~d$^{-1}$, $\beta$~Cephei ($\beta$~Cep) stars are main-sequence late O- and early B-type stars that have low-radial order p- and g-mode frequencies between about 2 and 25~d$^{-1}$ (see \citealt{bowman2020}). 
%There are also hybrid stars which present simultaneously g- and p-mode frequencies. 
However, we note that it is difficult to distinguish all the potential variability mechanisms operating in a massive star from its light curve alone.  When classifying all stars based on their TESS light curves and corresponding frequency spectra, we note that frequencies below about 0.5~d$^{-1}$ could be instrumental (e.g. imperfect detrending) or astrophysical (e.g. rotational modulation).

Each sector of available TESS data for a star is considered individually to check for the existence and consistency of variability seen across all its light curves. This is useful for a number of reasons; for example, to distinguish between SPB pulsation modes and SLF variability in the low frequency regime, as the g~modes in an SPB star are typically coherent and should be generally consistent across sectors, whereas SLF variability is non-periodic. 
%In most cases a target is classified as having only one, but sometimes two, types of pulsational variability. 
Some stars presented a combination of these features, in which cases all types of variability were taken note of. For classification purposes, priority is given to the dominant type of variability that is consistently observed across all sectors of available TESS data. 
%However all types of variability are taken note of for statistical considerations.
% which symbol is shown in Fig.~\ref{fig:sHRD}), 
% From this Fourier analysis, we categorised the targets by type, depending on the frequencies and stellar parameters we had determined. Stars showing coherent frequencies with $f \gtrsim 4.0$ d$^{-1}$, attributed to $p$-mode pulsations, and spectral types equal to or earlier than B3 were classified as $\beta$\,Cephei ($\beta$\,Cep) stars. Those with coherent frequencies between $1.0 \lesssim f \lesssim 4.0$ d$^{-1}$, attributed to $g$-mode pulsations, and spectral types later than B3 were classified as slowly pulsating B-type (SPB) stars. Finally, stars showing an excess of power in the frequency range $f \lesssim 1.5$ d$^{-1}$ that is not attributed to coherent $g$-mode pulsations, were classified as having stochastic low-frequency (SLF) variability. 
%noted for classification.

%\section{Method}

\section{Ensemble results}
\label{sec:distribution}

%{\onecolumn
\begin{table*}
\caption{The different variability types classified in our sample, which have been divided into subgroups of O- and B-type populations based on their SIMBAD spectral type. Column 4 denotes the number of stars that fulfil both binarity criteria from our RV calculations making them candidate spectroscopic binaries, column 5 denotes the number of SB2 systems, column 6 denotes the number of stars known in the SBX catalogue \citep{pourbaix2004,merle2026}, and the numbers of EBs are given in column 7, with the parentheses denoting those that were previously known in the literature.
% Columns two and three separate the populations into giants (I, II, or III) and dwarfs (IV and V). Columns four, five, and six correspond respectively to the stars fulfilling both criteria for binarity (BC), those that are included in the SBX catalog \citep{pourbaix2004}, and stars classified as EBs, with those in parentheses being previously known in the literature \citep{}. Column seven displays the distribution of stars showing rotational modulation in photometry as an indicator for magnetism, while column eight features targets that demonstrate the qualities of a candidate mCP. Columns nine and ten show the stars in the same containing emission features in their HERMES spectra, either as OBe stars or as BSGs. Finally, columns eleven, twelve, and thirteen break down the distribution of pulsators contained within the sample. 
The full table of stellar parameters and classifications are available in the supplementary material.}
%\resizebox{\textwidth}{!}{%
% \begin{tabular}{c|cc|cccc|cc|cc|ccc|c}
\begin{tabular}{c|cc|cccc|ccc|ccc|c}
\hline \hline
 % & \multicolumn{2}{c}{Luminosity class} & \multicolumn{4}{c}{Binarity} & %\multicolumn{2}{c}{Magnetic candidates} 
 % \multicolumn{3}{c}{Other} & \multicolumn{3}{c}{Pulsation} &  \\
% SpT & Giant & Dwarf & RV criteria & SB2 & SBX & EB & RotMod & %CP & 
% OBe & BSG & SPB & $\beta$\,Cep & SLF & Total \\ \hline
% O & 28 & 33 & 2 & 2 & 2 & 4 (0) & 5 & %0 & 
% 9 & 4 & 0 & 7 & 48 & 61 \\
% B & 471 & 341 & 49 & 18 & 26 & 63 (34) & 143 & %70 & 
% 118 & 22 & 197 & 244 & 220 & 812 \\
 & \multicolumn{2}{c}{Luminosity class} & \multicolumn{4}{c}{Binarity} & %\multicolumn{2}{c}{Magnetic candidates} 
 \multicolumn{3}{c}{Pulsation} & \multicolumn{3}{c}{Other} &  \\
SpT & Giant & Dwarf & RV criteria & SB2 & SBX catalogue & EB & SPB & $\beta$\,Cep & SLF & RotMod & OBe & BSG & Total \\ \hline
O & 28 & 33 & 2 & 2 & 2 & 4 (0) & 0 & 7 & 48 & 5 & 9 & 5 & 61 \\
B & 471 & 341 & 49 & 20 & 26 & 63 (37) & 197 & 244 & 220 & 143 & 122 & 28 & 812 \\
\hline \hline
\end{tabular}
%}
\label{tab:statistics}
\end{table*}%}

Of the 873 stars included in our sample, based on our analysis of TESS mission data we have classified based on frequency analysis 197 stars with %high-radial order 
g-mode pulsations characteristic of SPB stars, 251 stars with %low-radial 
p- and/or g~modes typical of $\beta$\,Cep stars, and 268 stars showing SLF variability (see Section~\ref{sec:pulsations}). This represents a pulsator fraction of over 80~per~cent, which is consistent with previous studies \citep{bowman2019, burssens2020}. We also identify 67 eclipsing binaries (EBs), 30 of which were not previously known in the literature. The TESS light curves have also yielded 148 stars to show rotational modulation (sec Section~\ref{sec:magnetism}). From analysis of the available HERMES spectra, we identify 353 of the 873 stars to have RV variability (or 76 per~cent of the 464 stars with at least 2 spectra), with 51 of these satisfying the two statistical criteria for binarity (see Section~\ref{sec:radvel}). 
%, and 93 candidate-magnetic CP stars (sec Section~\ref{sec:magnetism}). 
Finally, we also identify 22 SB2 systems, 131 OBe stars, and 33 \revision{BSGs}, %\textbf{and 93 candidate-magnetic CP stars,} 
all of which we do not investigate further in this study and leave for future work since such systems require dedicated individual analysis. The full breakdown of all our classifications is provided in Table~\ref{tab:statistics}. 

\subsection{Spectroscopic Hertzsprung--Russell (HR) diagram}
\label{sec:sHRD}

\begin{figure*}
    \centering
    \includegraphics[width=2\columnwidth]{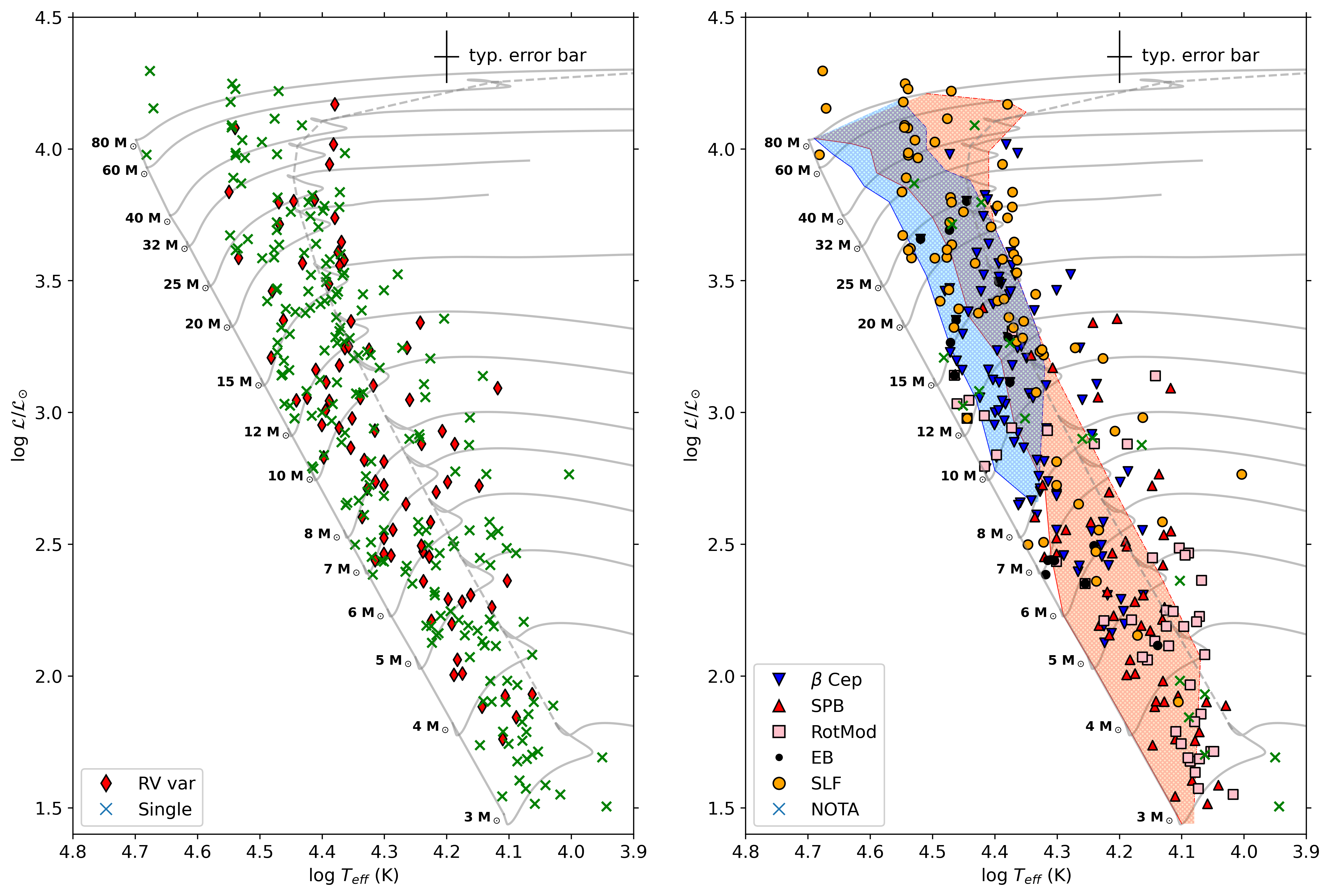}
    \caption{Spectroscopic HR~diagrams for 322 stars in our sample with available stellar parameters from the \emph{Gaia} ESP-HS database, which show our classifications based on HERMES spectroscopy (left) and TESS photometry (right). For both cases, we use the evolutionary tracks for a range of masses (solid grey lines) from \citet{burssens2020}. The zero-age main sequence (ZAMS) and terminal-age main sequence (TAMS) are denoted by solid and dashed grey lines, respectively. For the photometric case in the right panel, the theoretical instability strips corresponding to p-mode (blue) and g-mode pulsations (red) from \citet{burssens2020} are also shown. Stars not presenting any of the other features shown in the respective legends are labelled as `NOTA' (None Of The Above).}
    \label{fig:sHRD}
\end{figure*}

Once satisfied with the visual classification and corresponding stellar parameter determination, we generated the spectroscopic HR~diagrams shown in Fig.~\ref{fig:sHRD}%. 
%In Fig.~\ref{fig:sHRD}, we show spectroscopic HR~diagrams of the sample, 
, in which the spectroscopic luminosity values are defined as $\mathcal{L}~=~T_{\text{eff}}^4/g$ following \citet{langer2014}, for the 322 stars with available parameters from \emph{Gaia} ESP-HS database. We note that only 37~per~cent of our sample have such parameters. 
% In Fig.~\ref{fig:sHRD}, we show spectroscopic HR~diagrams of the sample, in which the spectroscopic luminosity values are defined as $\mathcal{L}~=~T_{\text{eff}}^4/g$ following \citet{langer2014}. The right-hand side shows the 322 stars with available parameters from \emph{Gaia} ESP-HS database. We note that this is the case for only 37~per~cent of our sample. As such, on the left-hand side, we complement the figure with the targets that do not have \emph{Gaia} ESP-HS parameters, using their calculated TLUSTY parameters instead, to demonstrate that these two parameter spaces are consistent with one-another.

We choose to use the \emph{Gaia} ESP-HS parameters, instead of the TLUSTY-derived parameters purely when plotting an HR~diagram, because the discrete sampling of the TLUSTY grid does not yield as informative a visual distributions in the HR~diagram. For the remainder of this article, however, we rely exclusively on the T$_{\rm eff}$, $\log\,g$, and $v\,\sin\,i$ parameters we determine from synthetic spectrum fitting using the TLUSTY grids. In the left panel of Fig.~\ref{fig:sHRD} we have denoted the subsamples of 
% stars with emission lines in at least one spectroscopic epoch (i.e. OBe stars), 
stars with significant RV variability, compared to the population without.
% 353 stars with significant RV variability (i.e. RV var), 93 stars with the continuum depression near 5200 \AA{} region and strong Si\,II lines at 5041 and 5055 \AA{} as candidate-magnetic chemically peculiar (i.e. mCP), 
% % stars with spectra characteristic of blue supergiant stars (i.e. BSG), 
% 75 candidate mCP stars with RV variability (i.e. RV + CP), and finally 382 OB stars with none of the above (i.e. NOTA).} \textcolor{red}{(Don't consider `Insufficient data' targets for this number?)

In Fig.~\ref{fig:sHRD}, we also include non-rotating evolutionary tracks calculated by \citet{burssens2020} with the {\sc MESA} stellar structure and evolution software package \citep{paxton2011, paxton2013, paxton2015, paxton2018, paxton2019}. The evolutionary tracks are for solar metallicity ($Z_{\text{ini}} = 0.014$), a heavy element mixture appropriate for massive stars from \citet{nieva2012}, OP opacity tables \citep{paxton2011}, a scaled mass-loss rate with a factor of $0.5$ compared to \citet{vink2001}, an exponential diffusive convective boundary mixing (CBM) prescription of $f_{\text{CBM}} = 0.02$, and a constant minimum envelope mixing value of $\log D_{\text{env}} = 1.0$ \citep{burssens2020}. The evolutionary tracks cover birth masses between 3 and 80~M$_\odot$, which is sufficient to cover our sample, and are calculated from the zero-age main sequence (ZAMS) up to near-depletion of core hydrogen mass fraction. In addition, from these models we define the terminal-age main-sequence (TAMS) as the point in the models where $X_c < 10^{-5}$. 
%The ZAMS and TAMS are shown as the solid and dashed grey lines in Fig.~\ref{fig:sHRD}. 
We emphasize, however, that the evolutionary tracks shown do not include binarity, magnetism, are non-rotating, and only include a single prescription for envelope and \revision{CBM}, all of which are known to play important roles in stellar evolution \citep[e.g.][]{ekstrom2012, Aerts2019, temaj2024}.

In the right panel of Fig.~\ref{fig:sHRD}, we show the same sample but with symbols that denote the dominant photometric variability as determined by our analysis of the TESS light curves and corresponding frequency spectra. 
These include $\beta$~Cep stars, SPB stars, stars with rotational modulation (RotMod), eclipsing binaries (EB), stars with stochastic low-frequency (SLF) variability (see \citealt{bowman2019b}), and any stars showing none of the above (i.e. NOTA).
% These include 251 $\beta$~Cep stars, 197 SPB stars, 148 stars with rotational modulation (RotMod), 68 eclipsing binaries (EB), 268 stars with stochastic low-frequency (SLF) variability (see \citealt{bowman2019b}), and 174 stars showing none of the above (i.e. NOTA).
% Pulsating stars that are also EBs will show both symbols. 
In addition, we have also included the theoretical pulsation instability regions calculated by \citet{burssens2020} by solving the non-adiabatic stellar pulsations equations using GYRE \citep{townsend2013, townsend2018}. The two instability regions denote where one expects to find low-radial order p~modes (blue; $1 \leq n \leq 5$), and a mix of low- and high-radial order g~modes (red; $-50 \leq n \leq -1$), for angular degrees of $0 \leq \ell \leq 2$ and $1 \leq \ell \leq 2$ for p~modes and g~modes, respectively. The hotter boundaries of the p- and g-mode instability regions are denoted by solid lines in Fig.~\ref{fig:sHRD}, whereas the cooler boundaries are denoted by dashed lines. The {\sc GYRE} calculations of \citet{burssens2020} only calculated instability regions for main-sequence {\sc MESA} models, and therefore did not include the post-main sequence phase of stellar evolution. 

We emphasize that any calculated instability region is only applicable to the specific evolutionary tracks used as input. This means that any difference in rotation rate, interior mixing, or mass loss rate, which inevitably changes the evolutionary tracks, would also change the location of the instability regions. Moreover, rotation not only changes the amount of mixing inside a star, but also influences the eigenfrequencies and the balance of driving and damping processes of pulsations (see discussions by \citealt{townsend2005, szewczuk2017}). 
%The blue edges of the instability strips shown in Fig.~\ref{fig:sHRD} are resolved by the calculations and thus are represented by a solid line, whereas the red edge is shown as a dashed line as the calculations are only performed up to core hydrogen depletion, and therefore doesn't consider later stages of evolution. 
Therefore, the evolution of stars, and whether they are expected to pulsate or not, within the mass regime discussed in this study are very sensitive to the choice of input parameters, such that small changes in parameter combinations can have large impact on the location  of both the evolutionary tracks and the corresponding instability regions (see also \citealt{paxton2015}). This means we consider the evolutionary tracks 
%and instability regions 
in Fig.~\ref{fig:sHRD} to be representative, but not perfect. Similarly, the instability region calculations only include a single excitation mechanism and are therefore not expected to give a complete image of where we can expect pulsations \citep[see, e.g.][]{hey2024}.

\subsection{Comparing parameters derived from TLUSTY and \emph{Gaia}}
\label{sec:TvG}

\begin{figure*}
    \centering
    \includegraphics[width=2\columnwidth]{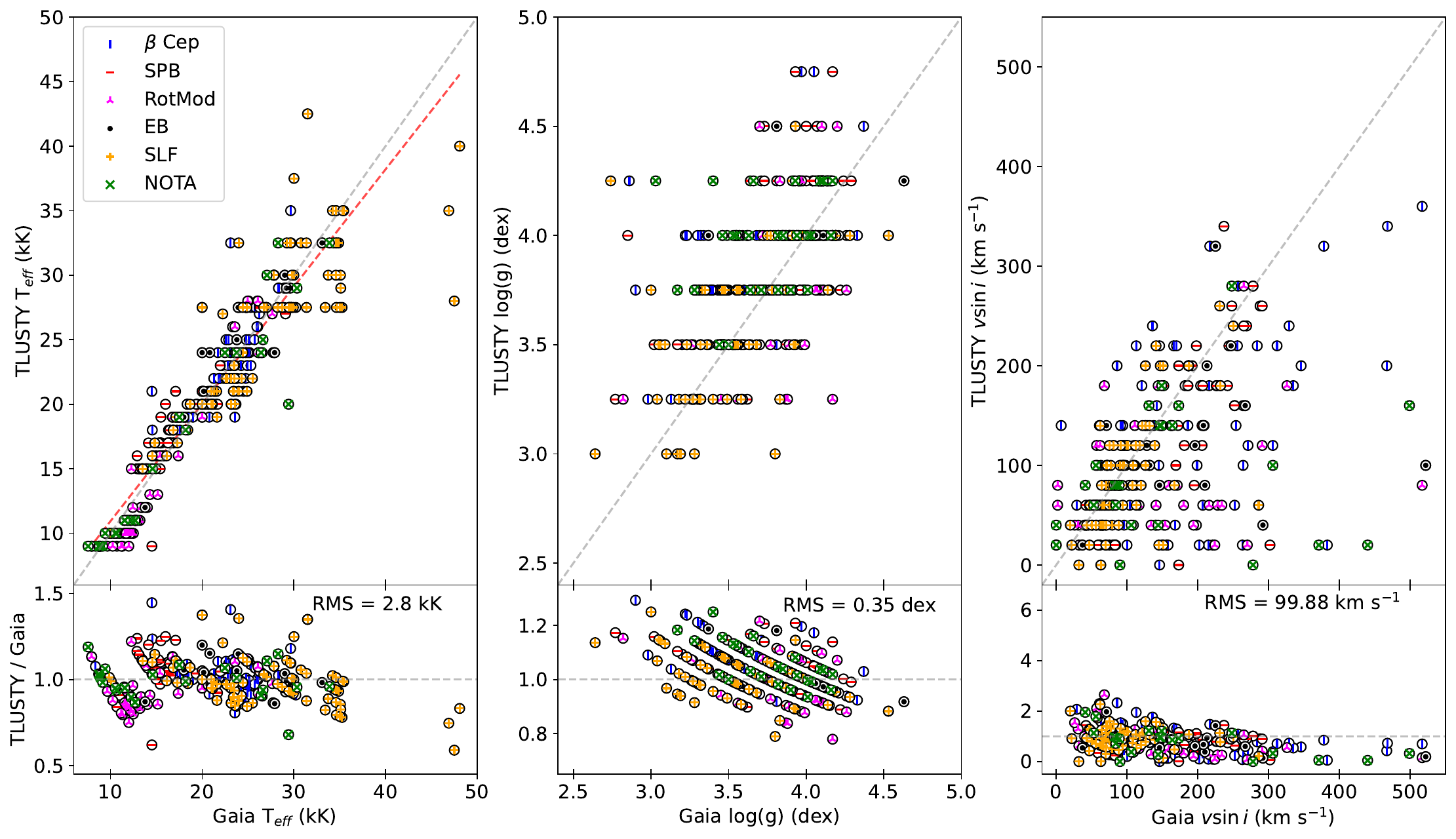}
    \caption{Comparisons of stellar parameters from the \emph{Gaia} ESP-HS database to those determined in this study using a grid-based fitting of TLUSTY atmospheric models. Targets have been labelled with respect to their photometric classification. Grey dashed lines corresponding to unity have been drawn for clarity, and an additional red-dashed line has been added for the T$_{\text{eff}}$ plot, representing a linear regression to the data points (see Section~\ref{sec:TvG}). The bottom row corresponds to the TLUSTY/\emph{Gaia} residuals for each parameter, and the resulting RMS values.}
    \label{fig:TvG}
\end{figure*}

In Fig.~\ref{fig:TvG}, we compare the atmospheric parameters deduced from our grid-based fitting of {\sc TLUSTY} models with those available from the the \emph{Gaia} ESP-HS database. Upon first inspection, while the two sets of $T_{\text{eff}}$ values appear to be largely consistent, there are a subset of stars which have fairly large differences.
% Specifically, the largest discrepancies are for $T_{\rm eff}$, $\log\,g$ and $v\,\sin\,i$ between {\sc TLUSTY} and \emph{Gaia} is for the OBe stars. 
The (dis)agreements in Fig.~\ref{fig:TvG} are somewhat worse for $\log\,g$ and $v\,\sin\,i$, and can be understood because of a number of factors, which we discuss below.

Firstly, \citet{fouesneau2023} suggests that $T_{\text{eff}}$ values determined from ESP-HS tend to be underestimated by about 2000~K for stars with $T_{\rm eff} \gtrsim 25~000$~K when compared to large spectroscopic surveys, such as DR6 of the Large Sky Area Multi-Object Fibre Spectroscopic Telescope \citep[LAMOST;][]{zhao2012,cui2012,xiang2022}. This is somewhat visible in the left panel of Fig.~\ref{fig:TvG}, where we see a slight bias of points above the unity line, further accentuated by the linear regression to the data points (red dashed line) becoming increasingly offset beyond about $\sim$25~000~K. While this offset does not appear to be systematically equivalent to 2000~K, the fit suggests that the offset deteriorates towards increasingly higher values of $T_{\rm eff}$.

Secondly, \emph{Gaia} $\log\,g$ values calculated by the astrophysical parameters inference system \citep[Apsis;][]{bailerjones2013} are determined primarily through fitting a grid of synthetic spectra in BP/RP photometry and low-resolution RVS spectra (when available) using a $\chi^2$ minimisation, using the Balmer and Paschen lines as anchors. These spectral lines reach a maximal strength at $T_{\rm eff} \simeq 8500$~K and, above this threshold, there is a noticeable underestimation of $\log\,g$ in the GSP-Phot values. This offset is somewhat corrected for in ESP-HS by using a grid of synthetic spectra tailored to hot stars and by focusing on the Balmer line series and helium lines for surface gravity determination. For example, an improvement in accuracy by a factor of two with respect to GSP-Phot is expected for the magnitude range of our sample \citep{fouesneau2023,gaia2023c}. That said, a systematic scatter of about 0.2 to 0.4~dex is expected for the derived  $\log\,g$ values from \emph{Gaia} ESP-HS, particularly when extending to the higher temperature ranges ($\gtrsim 30~000$K). This scatter is seen when comparing the \emph{Gaia} ESP-HS values to our grid-based TLUSTY fitting values in the middle panel of Fig.~\ref{fig:TvG}. We emphasize that the step size of $\log\,g$ in the TLUSTY grid is larger than that of \emph{Gaia} in a relative sense, which produces `rows' of values to stack on top of each other in the middle panel of Fig.~\ref{fig:TvG}. 

Finally, to measure $v\,\sin\,i$ in ESP-HS, \emph{Gaia} assumes that effectively almost all spectral line broadening is due to rotation, by convolving and fitting the full synthetic spectra with rotational broadening kernels \citep{fouesneau2023}, which is not provided at all by the standard GSP parametrisation. While this is a reasonable approximation for low- and intermediate-mass stars (see \citealt{GRAY_BOOK}), the spectral lines of massive stars have considerable additional broadening mechanisms. For example, microturbulence and macroturbulence both have a significant impact on the shape of spectral lines (see \citealt{aerts2009, simondiaz2017, kalita2025}). By default, \emph{Gaia} assumes a standard value of $v_{\rm mic} =$~2~km~s$^{-1}$ for microturbulence, which aligns with the available value in the B-star grid of {\sc TLUSTY} models we used in this work, though differs to the 10 km~s$^{-1}$ value utilised for the O-star {\sc TLUSTY} grid. Neither \emph{Gaia} nor the {\sc TLUSTY} models include macroturbulence as an additional fitting parameter, and are therefore not considered as a source of line broadening. Another factor could be that of additional SB2 systems that went unnoticed in our spectroscopic analysis, where blended lines might appear as one star in certain epochs and as two stars in another, resulting in an overestimation of the width of the given line in one instance or the other. We nonetheless observe that the majority of points (and in particular almost all of the outliers) fall below the unity line in the right panel of Fig.~\ref{fig:TvG}, indicating that \emph{Gaia} almost systematically overestimates the $v\,\sin\,i$ values with respect to those we calculated using the {\sc TLUSTY} models.

% The lack of macroturbulence as an additional fitting parameter is likely why the $v\,\sin\,i$ values determined by \emph{Gaia} are systematically higher with respect to those we calculated using the {\sc TLUSTY} models. This is supported by the right panel of Fig.~\ref{fig:TvG}, as the majority of points (and in particular almost all of the outliers) fall below the unity line. On the other hand, we also did not include macroturbulence as a broadening parameter in our grid-based fitting of {\sc TLUSTY} models. 

Therefore, we conclude that {\emph Gaia} ESP-HS produces reliable $T_{\rm eff}$ values, on average, for massive stars, which are useful for ensemble analysis. However, high-resolution spectroscopy is highly beneficial when studying $\log\,g$ and $v\,\sin\,i$ for massive stars, especially for those in binary systems (see Section~\ref{sec:radvel}).

%Something similar occurs for $\log(g)$, except that the values are instead underestimated by \emph{Gaia}.

% Returning to the subset of stars with the largest discrepancies, the OBe stars, unsurprisingly, these {\sc TLUSTY} models are not ideal for fitting emission features seen in OBe stars. Therefore, without manually masking emission lines before fitting with {\sc TLUSTY}, we obtain poor estimates for $T_{\rm eff}$, $\log\,g$, and $v\,\sin\,i$. While \emph{Gaia} likely suffers from similar issues, it benefits from having access to a larger temporal baseline of observations and several orders of magnitude more stars to determine mean stellar parameters, resulting in potentially more reasonable estimates. Performing in-depth synthetic spectrum fitting on a case-by-case basis for stars showing emission-line features is outside of the scope of this paper, but similar conclusions were found by \citet{nardini2025}. %As such, we elect to exclude these subsets of stars from any statistical comparisons based on stellar parameters performed in the remainder of this study.

\subsection{Population of evolved stars}

A minority of the stars in our sample appear to have evolved beyond the TAMS according to the evolutionary tracks calculated by \citet{burssens2020} and parameters determined from \emph{Gaia} ESP-HS, shown in Fig.~\ref{fig:sHRD}. This conclusion is based on evolutionary tracks that include only a modest amount of CBM, with the diffusive exponential prescription of $f_{\rm CBM} = 0.02$. Massive stars are expected to evolve rapidly across the Hertzsprung gap after the main sequence, so it may seem surprising to observe a significant portion of the sample located in this region of the HR~diagram. However, we note that the main-sequence lifetime of a massive star and thus the position of the TAMS are highly sensitive to the input parameters, for example rotation, mixing, and metallicity, used in the evolution tracks shown in Fig.~\ref{fig:sHRD} \citep[see][]{bowman2020,deburgos2025,martins2013,ekstrom2012,gilkis2021,castro2014}. As such, stars seemingly redder than the TAMS may still be main-sequence stars with large amounts of interior mixing, or undergoing blue loops, or the results of merger products, etc.

% On the other hand, the highest density of stars that are much cooler than the TAMS in Fig.~\ref{fig:sHRD} have masses between about 7 and 15~M$_{\odot}$. Their atmospheric parameters characterise them as blue supergiants (BSGs), and their TESS light curves reveal a diverse range of pulsations. This part of the HR~diagram is particularly interesting for asteroseismic studies, since such stars are rare and may be in the shell-hydrogen burning or core-helium burning phases of stellar evolution, and may be undergoing blue loops in their evolution (see \citealt{bowman2019, Georgy2021, Bellinger2024}). However, finding pulsating BSGs is rare and performing asteroseismic modelling of such stars is challenging (e.g. \citealt{Saio2006}).
%displayed in this region might appear more evolved than they would be if the models were fine-tuned to take into account the specific parameters of a given star. 
% \textcolor{red}{To assess this, we plot the empirical TAMS line defined by \citet{deburgos2025} for the luminosity range of our sample, described by}
% \begin{equation}
%     \log \left(\frac{L}{L_\odot}\right) = 4.7\times10^{-4}\;\text{T}_{\text{eff}} -5.42\;[\text{dex}].
% \end{equation}
% which becomes
% \begin{equation}
%     x=x
% \end{equation}
% when converting from bolometric to spectroscopic luminosity and using $\log\,g = 3.3$

\section{Pulsating stars}
\label{sec:pulsations}

%\subsection{Pulsator type classification}

\begin{figure*}
    \centering
    \includegraphics[width=0.98\linewidth]{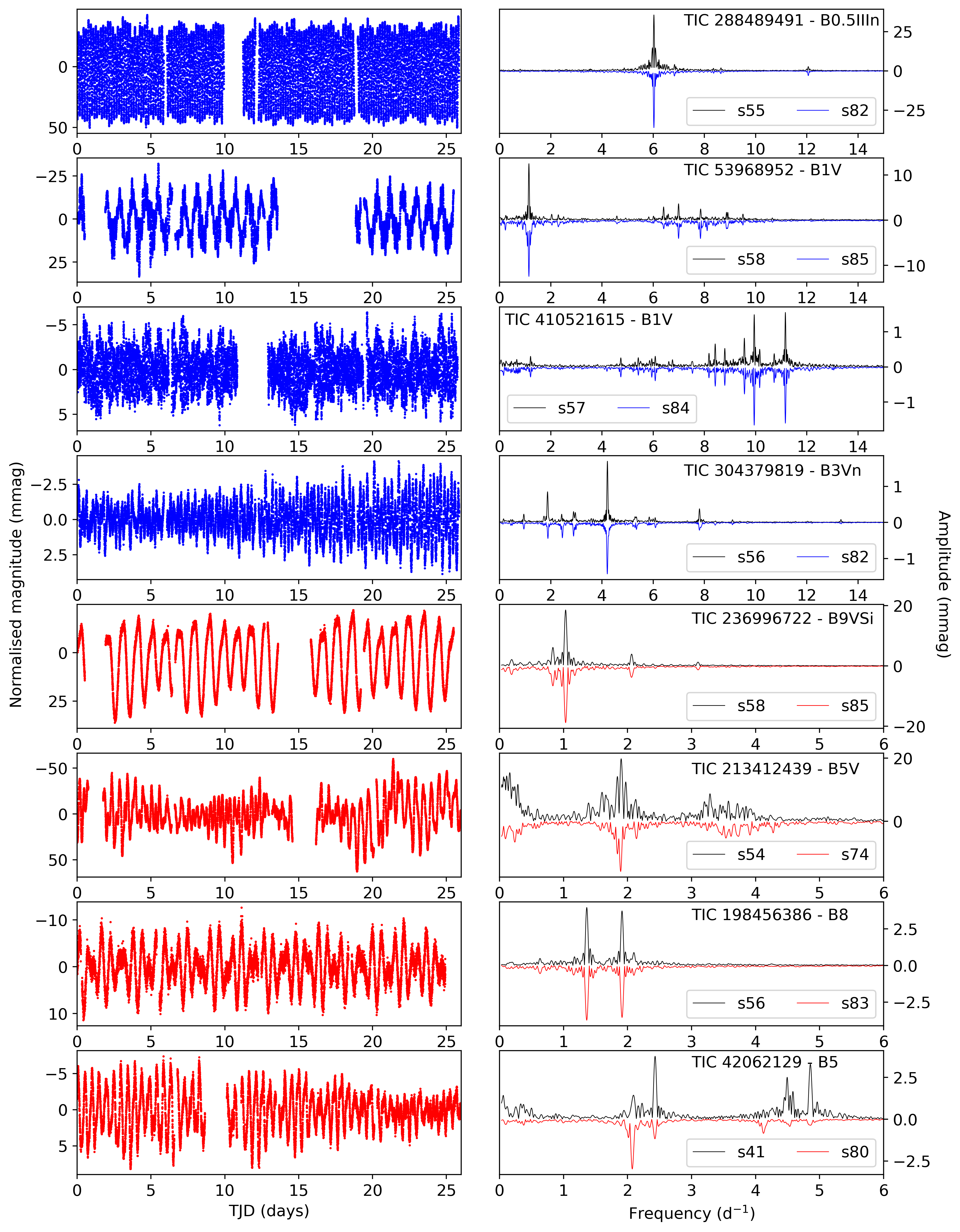}
    \caption{TESS light curves and Lomb-Scargle (LS) periodograms for four $\beta$~Cep (blue; TIC\,288489491, TIC\,53968952, TIC\,410521615 and TIC\,304379819) and four SPB stars (red; TIC\,236996722, TIC\,213412439, TIC\,198456386 and TIC\,42062129). The LS~periodograms show two different sectors in blue (red) and black for comparison purposes, whereas the light curves show only the TESS sector correspondingly coloured in the blue (red) periodograms.}
    \label{fig:pulsators}
\end{figure*}

Similar to the previous study of massive stars in the Southern hemisphere by \citet{burssens2020}, we find a high pulsator fraction ($>80$~per~cent) for galactic massive stars. Typical examples of the TESS light curves and corresponding frequency spectra from the coherent pulsator classes of $\beta$~Cep and SPB stars are shown in Fig.~\ref{fig:pulsators}. Within the sample, there exist many coherent pulsators that offer excellent prospects for forward asteroseismic modelling (e.g. \citealt{pedersen2021, burssens2023}), which is the subject of future work (Scott et al. in prep).

\subsection{Pulsation instability regions}

As shown in the right panel of Fig.~\ref{fig:sHRD}, we find that the majority of the $\beta$~Cep and SPB pulsators in our sample lie within their respective instability regions in the HR~diagram. On the other hand, there is a small subset of evolved $\beta$~Cep stars located cooler than the TAMS as calculated in the evolutionary tracks by \citet{burssens2020}. As mentioned previously, these evolutionary tracks are non-rotating and only assume a single set of physical assumptions, such that a larger amount of mixing would prolong a star's main-sequence lifetime, and hence displace the TAMS to cooler temperatures. Although the number of post-main sequence $\beta$~Cep stars is low in the literature, with some even being explicitly referred to as near the TAMS \citep[e.g. V453 Cyg A;][]{pavlovski2009,southworth2020}, the number of 
%post-main sequence 
$\beta$~Cep stars beyond the model grid TAMS observed in our sample only represent a small fraction of the full sample. They could potentially be explained by varying the input physics of our evolutionary tracks (i.e. mixing parameters or binary effects) to extend the main sequence to cooler effective temperatures to encapsulate all the observed $\beta$~Cep stars.

\begin{figure}
    \centering
    \includegraphics[width=0.95\linewidth]{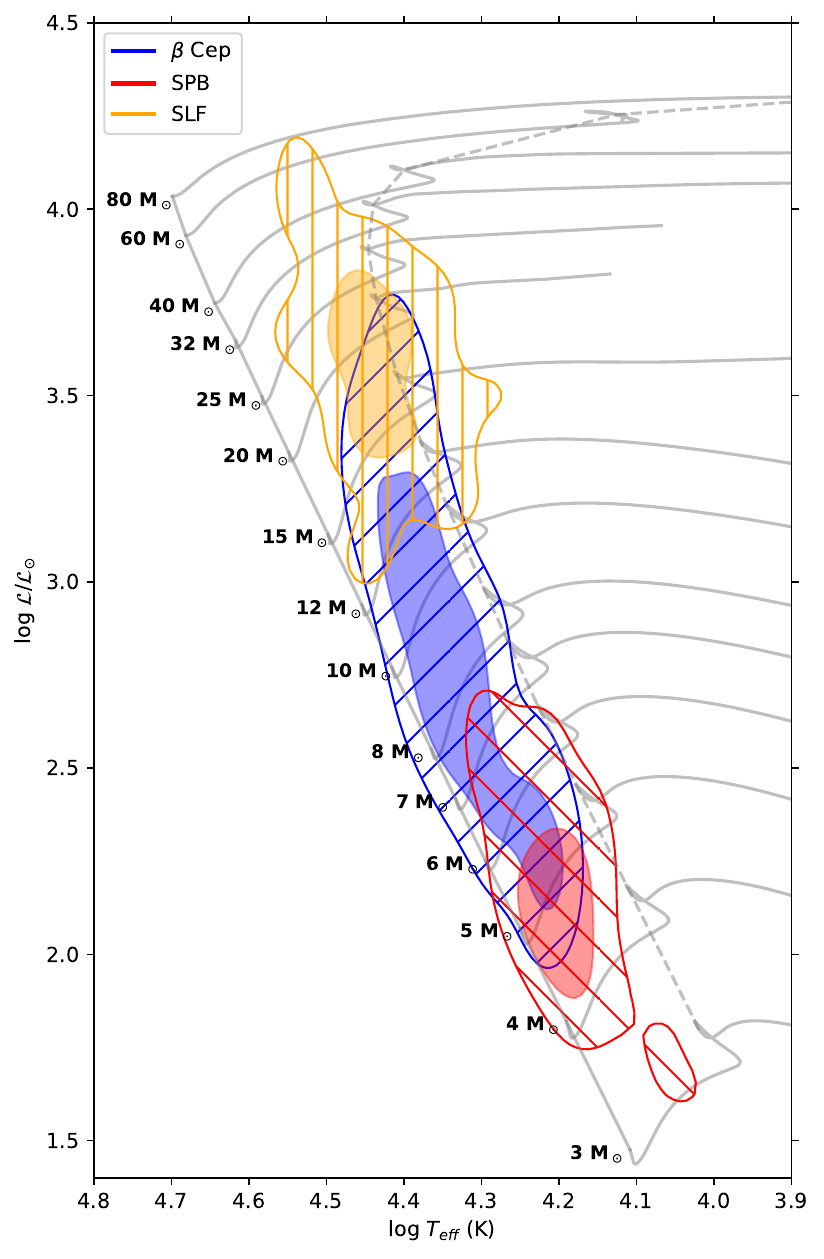}
    \caption{Spectroscopic HR~diagram showing the distribution density of the three main pulsator types found in the sample ($\beta$~Cep, SPB, and SLF). Contours are drawn at the 66th and 33rd percentiles in terms of population intervals for the hatched and filled regions, respectively. 
    %A dashed line connects the models with $X_{\rm c} < 10^{-5}$, to approximate the terminal-age main sequence.
    }
    \label{fig:contours}
\end{figure}

In the HR~diagram in Fig.~\ref{fig:contours}, we show density contours that are defined as large fractions of stars showing a common type of pulsation. Therefore, these contours represent the empirical instability regions for each of the three main pulsator types: $\beta$~Cep, SPB, and SLF variability. Specifically, the hatched contours represent the region containing 66~per~cent of each pulsator population, and the filled contour represents the 33rd~percentile of the observed population for each pulsator type.

For the $\beta$\,Cep stars, the sample of 119 stars from \citet{fritzewski2025} follow a similar distribution to those in our sample. However, while noting the difference between bolometric and spectroscopic luminosity, our sample is shifted towards lower masses, with at least 66~per~cent of the population being contained within the 5 to $20$~M$_\odot$ range (with a few outliers up to $60$~M$_\odot$ seen in Fig.~\ref{fig:sHRD}), compared to their range of reported evolutionary masses: $6.5-32$~M$_\odot$. A similar comparison of the location of the empirical SPB instability region is performed with the observed population of 52 SPB stars from \citet{pedersen2022}, which range from $3-8$~M$_\odot$. Our sample of SPB stars is mostly in agreement, with at least 66~per~cent of the population covering the same evolutionary mass range. Interestingly, despite the number of excited g~modes in SPB stars being generally larger for masses of about 3-4~M$_{\odot}$ (see figure 11 of \citealt{Pedersen2020}), the mass range in which the highest fraction of SPB stars is found in our sample is between 4-6~M$_{\odot}$. We postulate that this in predominately an impact of metallicity, for example, the difference between OP and OPAL opacity tables used by us and \citet{Pedersen2020} respectively (see \citealt{paxton2015} and \citealt{Moravveji2016b}). Finally, \citet{bowman2020b} find that majority of main-sequence stars with SLF variability as their dominant variability type are O-type stars, which is in agreement with our sample. 

More generally, the two coherent pulsator types, $\beta$~Cep and SPB stars, are found across the whole main sequence in Fig.~\ref{fig:contours}, which contains evolutionary tracks with an assumed amount of CBM being $f_{\rm CBM} = 0.02$ \citep{burssens2020}. This lends support to the idea that this amount of CBM is reasonable for a population study of main-sequence stars with masses between about 3 and 15~M$_{\odot}$. However, the population of stars with SLF variability are mostly found towards the latter half of the main sequence and are generally found near the TAMS. For masses above $\sim$15~M$_{\odot}$, assuming all stars with SLF variability are main-sequence stars in our study, a larger amount of CBM than $f_{\rm CBM} = 0.02$ (or the inclusion of rapid rotation in the models) would be needed to shift the TAMS to cooler temperatures and encapsulate these stars as main-sequence stars. This is, of course, heavily dependent on the various parameters that dictate the location of the TAMS, as described in Sec.~\ref{sec:sHRD}.

There exists a subset of $\beta$\,Cep stars in our sample with masses between 5--7~M$_\odot$, determined by their position in the HR~diagram, which is atypically low mass for such stars (supported by the findings of \citealt{fritzewski2025}). These stars are approaching the regime where we might expect to find $\delta$\,Scuti stars, which also present high-frequency pulsation peaks (between $\sim5-50$~d$^{-1}$), though in these stars it is due to the He\,{\sc ii} opacity bump as opposed to the metal opacity bump in the case of $\beta$\,Cep stars. Indeed, particularly in the case of strong magnetism, it has been found that the $\delta$\,Scuti instability strip can be extended to higher temperatures and luminosities \citep{thomson2025}. Distinguishing these two populations would require additional photometric follow-up and investigation, using asteroseismic modelling to determine the nature of the excitation mechanism.
% \textcolor{red}{We also observe a population of cooler stars showing SLF variability, located in an overlapping region between the SPB and $\beta$\,Cep contours, though these could prove to be unresolved $g$-modes that are indistinguishable from SLF variability in only 1 sector of TESS data.} 

Finally, we note that there is significant overlap among the three different pulsator types in Fig.~\ref{fig:contours}. This is not unexpected since so-called `hybrid' pulsators of $\beta$~Cep and SPB stars have been known for some time (see \citealt{Handler2009}). Moreover, many stars show multiple distinct variability types in our sample; for example, both high-frequency p~modes typical of $\beta$~Cep stars as well as SLF variability. In this exercise, however, we have only labelled the dominant variability type of each star, which means that hybrid stars are not a distinct pulsator type.

% \begin{figure}
%     \centering
%     \includegraphics[width=0.95\linewidth]{Figs/HRD_contours_v2.pdf}
%     \caption{50\% criterion.}
%     \label{fig:placeholder}
% \end{figure}

\subsection{SLF variability}

The dominant physical cause of SLF variability in massive stars remains debated in the literature, and may even arise from a combination of different physical mechanisms that dominate in different parts of the HR~diagram (see \citealt{bowman2023}). One proposed mechanism is internal gravity waves (IGWs) excited at the turbulent convective core boundary \citep{rogers2013, rogers2017,edelmann2019,anders2023,thompson2024,pathak2025}, which propagate to the stellar surface and produce a broad, low-frequency power excess in photometry \citep{blomme2011,aerts2015,bowman2019, bowman2020b}. Another contribution to SLF variability can come from the consequences of sub-surface convection zones which are associated with opacity bumps, typically corresponding with the H, He or Fe ionisation regions/opacity peaks. These zones can excite IGWs or induce turbulence that manifests as SLF variability \citep{cantiello2021,schultz2022}. A third mechanism is variability in the radiatively driven winds of massive stars, for which stochastic changes in wind density, clumping, or large-scale structures introduce low-frequency brightness fluctuations, which is particularly important for the most luminous and evolved massive stars \citep{krticka2018,krticka2021}. The relative importance of these mechanisms is expected to depend on stellar mass, evolutionary state, and wind strength.

SLF variability has emerged as a ubiquitous phenomenon in massive stars with the advent of high-precision space photometry \citep{bowman2019}. The morphology of SLF variability correlates with mass and evolutionary state, demonstrating a direct link between SLF variability and a star's internal stellar structure \citep{bowman2020b, bowman2022b, pedersen2025}. More recent studies have shown that the observed SLF properties are largely insensitive to metallicity \citep[][Van Daele et al., submitted to MNRAS]{bowman2024}, which supports the interpretation that the dominant driving mechanism is likely not dominated by instabilities that depend directly on opacity (e.g. sub-surface convection). 

Whilst remaining agnostic on the cause(s) of SLF variability in our sample of massive stars, we find it to be ubiquitous in stars with masses above about $M \gtrsim 12$~M$_{\odot}$, with a handful of suggested cases appearing in masses as low as 4~M$_{\odot}$, and it is generally the dominant form of photometric variability in stars above about $M \gtrsim 15~M_{\odot}$, as shown in Fig.~\ref{fig:contours}. This is in agreement with previous studies of galactic massive stars \citep{bowman2019,bowman2019b,bowman2020b}. In a future study, we shall investigate the sub-sample of stars with SLF variability in detail (Van Daele et al., in prep.). 
% While it appears that some SLF variability is observed between 6 and 8~$M_{\odot}$ near the TAMS in our sample, it is likely that this originates from unresolved g~modes in SPB stars with only short TESS light curves available. Since the g-mode density in the SPB frequency regime is notably higher for stars near the TAMS as compared to younger stars (see \citealt{buysschaert2018b}), these high-amplitude low-frequency signals can appear to look like evolved SPB pulsation modes rather than SLF variability, leading to a bias in our classification. 

\subsection{Pulsations and rotation}

\begin{figure}
    \centering
    \includegraphics[width=0.95\linewidth]{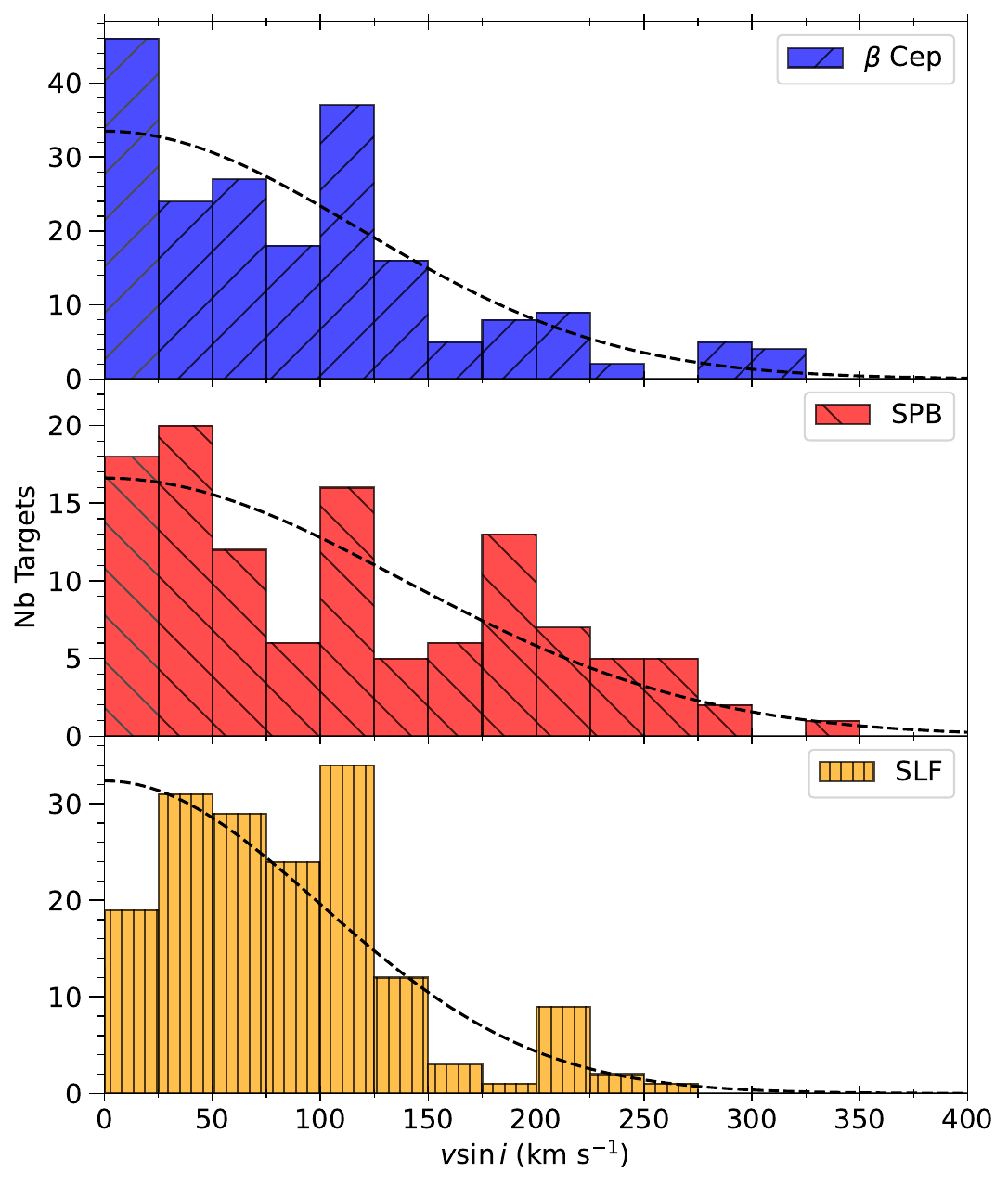}
    \caption{Distributions of $v\,\sin\,i$ values derived in this work from grid-based fitting of rotationally broadened TLUSTY atmospheric models for the three main pulsator types: 200 $\beta$~Cep stars, 116 SPB stars, and 167 stars with SLF variability as their dominant type of photometric variability. A normal distribution fit centred on 0 km~s$^{-1}$ (black dashed line) has been overplot for each subset. }
    \label{fig:vsini}
\end{figure}

The distributions of $v\,\sin\,i$ values we derived from grid-based fitting of rotationally broadened TLUSTY atmospheric models for each of the three main pulsator types are shown in Fig.~\ref{fig:vsini}. About half of the 664 pulsating stars are identified as slow-to-moderate rotators based on their projected surface rotational velocities. For example, 301 of the 664 pulsators have $v\,\sin\,i \leq 100$~km\,s$^{-1}$, and a further 134 stars have $100 < v\,\sin\,i \leq 200$ km\,s$^{-1}$. Additionally, 74 stars have $v\,\sin\,i \leq 30$~km\,s$^{-1}$, which may constitute a sample of very slow rotators. These distributions are consistent with the results of previous ensemble studies \citep[e.g.][]{simondiaz2014, simondiaz2017, burssens2020}. The few examples with values $v\,\sin\,i > 300$~km\,s$^{-1}$ have all been assessed manually to verify the goodness-of-fit of the TLUSTY model from which the parameter determination was performed, and are confirmed to be reasonable. As a reminder, we have omitted the OBe stars from our sample, so the presence of rapidly rotating stars without emission is interesting to follow up. For example, such rapidly rotating stars may be the products of binary interaction \citep[see, e.g.][]{demink2013} or OBe stars in quiescent periods.

The 200 $\beta$~Cep stars show a large range of $v\,\sin\,i$ values, with the majority being slow-to-moderate rotators ($0 < v\,\sin\,i \leq 150$~km\,s$^{-1}$). However, a fast rotating tail of the distribution reaches an upper value of about 300 km\,s$^{-1}$. The average rotation rate is 95~km\,s$^{-1}$. This is consistent with the findings of \citet{stankov2005}, who suggest that this might be due to a selection effect, as slower rotating $\beta$~Cep stars show the highest amplitude and most easily identified as pulsators. However, we note that the ensemble study of \citet{stankov2005} is based on ground-based detections of $\beta$~Cep stars, which have a typical photometric precision of order 1~mmag. Whereas space photometry, such as the TESS light curves we have used in the current study, have a typical photometric precision of about 10~$\mu$mag. 
%Moreover, the Nyquist frequency of our 2-min cadence TESS light curves means there is negligible amplitude suppression for higher frequencies (see \citealt{Bowman2017}, chapter 2). 
Therefore, our modern sample of $\beta$~Cep stars should be less biased towards finding faster rotators than \citet{stankov2005}, and yet we find similar results. Therefore, it is reasonable to conclude that $\beta$~Cep stars are generally slow-to-moderate in their surface rotation rates.

We see a similar distribution of $v\,\sin\,i$ values in the SPB stars to the $\beta$~Cep stars, though the range extends to slightly larger values of 350~km\,s$^{-1}$. The average rotation rate of the 116 SPB stars in our sample is 116~km\,s$^{-1}$.
%however there appears to be an under-representation of stars with $100 < v\,\sin\,i \leq 150$ km\,s$^{-1}$ that does not appear 
Finally, for the 167 stars with dominant SLF variability, the $v\,\sin\,i$ values are typically lower with the gross majority smaller than 150 km\,s$^{-1}$, with an average value of 89~km\,s$^{-1}$. %follow a %Gaussian-like 
%pseudo-normal distribution centred around $\sim$75 km\,s$^{-1}$, with a small tail towards higher values.

\section{Binarity}
\label{sec:radvel}

%Binarity strongly influences stellar evolution through a variety of processes, and appears to be ubiquitous amongst massive star populations.
%such as tidal interactions, mass transfer, angular momentum exchange, and mergers, all of which can significantly alter a star’s internal structure, surface composition, rotation rate, and evolutionary pathway (see \citealt{langer2012, marchant2024} for reviews). 
%Previous studies have shown that over 70~per~cent of massive stars are expected to interact with a companion at some point during their lifetime (e.g. \citealt{sana2012,demink2013}). 
To assess the fraction of our sample that are in multiple star systems, it is important to separate the binaries from other types of variability that may mimic the signatures of binarity \citep[e.g. pulsation, rotational modulation;][]{ijspeert2021, prsa2022}. 
%martins2017, mahy2020
The combination of spectroscopic and photometric data has revealed a variety of different types of binary systems within our sample, including many that also pulsate. %Specifically, our sample contains 56 eclipsing binaries. %and 10 spectroscopic binaries. 

\subsection{Eclipsing binaries}

From the analysis of the TESS photometry, we identify 67 eclipsing binaries. This includes 37 that have already been identified in the literature from previous surveys (e.g. \citealt{malkov2006,avvakumova2013,ijspeert2021,prsa2022,mowlavi2023, eze2024}). Therefore, we add 30 new eclipsing binaries in this work. Light curves for those with 2-min cadence TESS data are visible in Figs.~\ref{fig:EB_1}-\ref{fig:EB_5}, along with orbital period estimates. Of these 30, a half (16) have orbital periods shorter than 3~d. A few targets (e.g. TIC\,106781544) have large and unconstrained orbital periods ($P_{\text{orb}} \gtrsim 13.5$~d) because only a single eclipse is visible in a single TESS sector, thus making accurate period determination impossible. Unless those targets also had back-to-back sectors that contained at least one eclipse each, we could only determine a lower threshold for their orbital period.

We additionally find that 12 of the full list of 67 eclipsing binaries have $\beta$~Cep or SPB pulsations, as well as a further 11 systems with SLF variability. Both TIC\,91111448 \citep[][]{chen2022} and TIC\,434723918 \citep[][]{southworth2022} were previously known in the literature as pulsating ($\beta$~Cep) eclipsing systems (see also \citealt{eze2024}). Six of these EBs are also identified as SB2s in our spectroscopic analysis, which are particularly powerful laboratories for constraining stellar structure and evolution theory (see \citealt{southworth2025}). We leave the detailed analysis of the new (pulsating) EBs discovered in this work for a future study.

\subsection{Spectroscopic binaries}

Our sample contains a grand total of 124 binary systems, which includes 22 targets classified as SB2, 28 targets included in the SBX binarity catalogue \citep{pourbaix2004, merle2026}, 67 EB systems (of which 10 fulfil both binarity criteria), and 51 targets that fulfil both RV variability criteria described above (assuming $C = 20$~km\,s$^{-1}$). Considering exclusively binaries detected through our analysis of the HERMES spectra results in a total of 82 systems. %\revision{The significance of this result in discussed at length in Sec.~\ref{sec:binary-frac}.}

\begin{figure}
    \centering
    \includegraphics[width=0.95\linewidth]{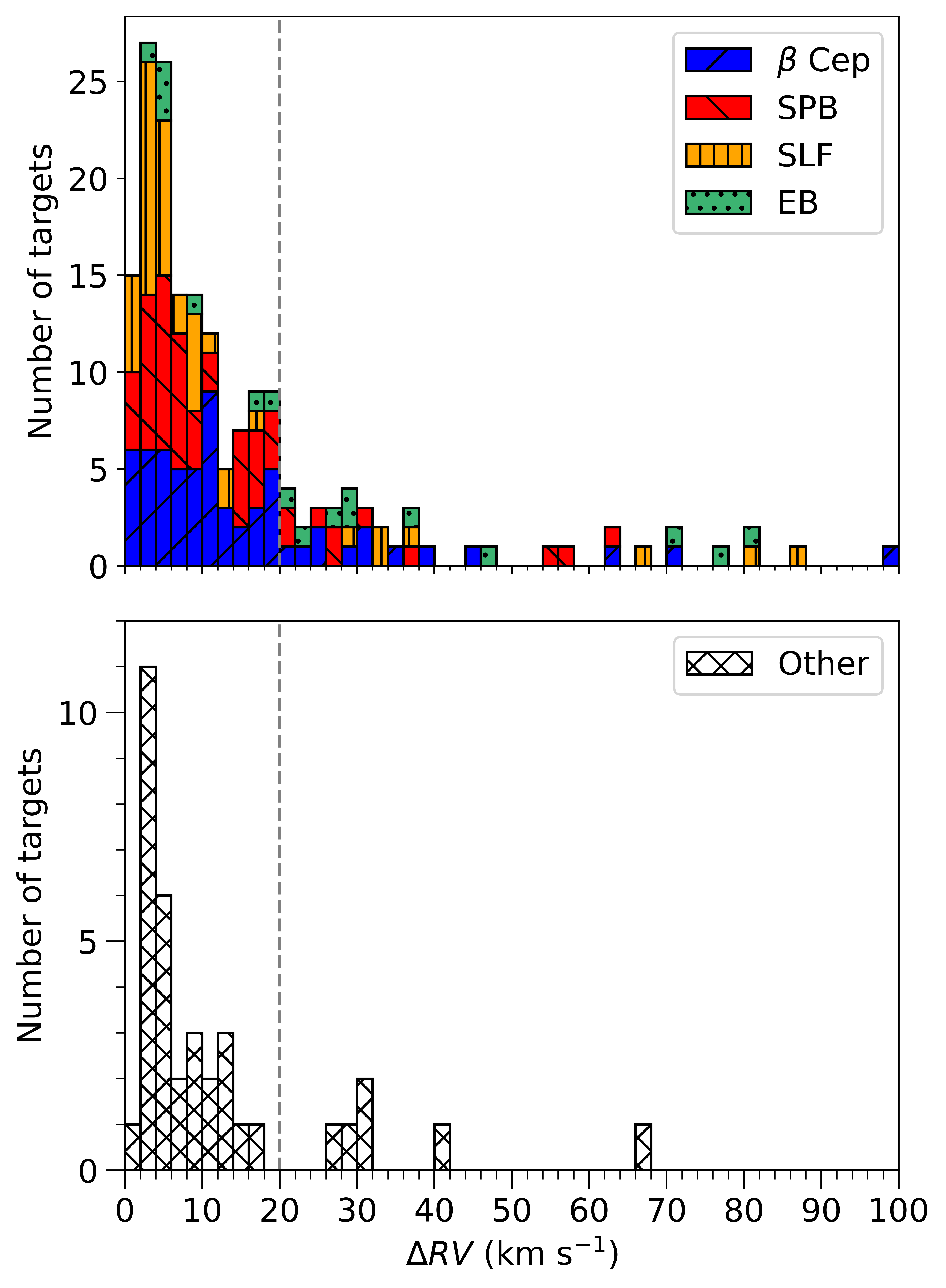}
    \caption{Top panel: Stacked distributions of the calculated $\Delta$RV values for the three main types of pulsating star in the sample, in comparison to the EB population. Bottom panel: $\Delta$RV value distribution for all targets in the sample that are not identified as a pulsator nor an EB from their TESS light curve. In both panels, the grey dashed line corresponds to the commonly adopted threshold for binarity of $C = 20$~km\,s$^{-1}$.}
    \label{fig:RVdist}
\end{figure}

In Fig.~\ref{fig:RVdist}, we show the $\Delta$RV distribution for the subset of pulsators that fulfil only the first binary detection criterion (cf. Eqn.~\ref{eq:rv1}), with a dashed grey line demonstrating the classically adopted $C=20$~km~s$^{-1}$ threshold for the second criterion (cf. Eqn.~\ref{eq:rv2}). Also in Fig.~\ref{fig:RVdist}, we have included the photometric classification of the stars based on the TESS light curves (see also Fig.~\ref{fig:sHRD}) to gain a perspective on the relationship between RV variability and the different pulsator types. Clearly, all three types of pulsator exist above and below the threshold of 20~km\,s$^{-1}$, which makes the value of $C$ in Eqn.~(\ref{eq:rv2}) a somewhat ambiguous choice. We have also included in Fig.~\ref{fig:RVdist} 18 EBs from our sample that satisfied the first binarity criterion (i.e. $\Delta RV > 4.0$), of which 12 have at least three spectra, to demonstrate that such systems also exist above and below the 20~km\,s$^{-1}$ threshold. %\revision{We discuss the choice of the 20~km\,s$^{-1}$ threshold in more detail in Sec.~\ref{sec:threshold}}.

%Importantly, there are a number of pulsating stars with very large $\Delta$RV values. 

\subsection{Discussion}

\subsubsection{Binary fraction}
\label{sec:binary-frac}

Based on a sample of 377 O- and B-type stars with two or more spectroscopic epochs (and no emission or SB2 features), about 22~per~cent \revision{(82/377)} of the stars have significant RV variability that passed both binary threshold criteria. \revision{This is much lower than previous studies \citep{sana2012,bodensteiner2021,banyard2022,mahy2022,nardini2025,frost2025}, who generally find observed binary fractions of order 50~per~cent or higher from dedicated spectroscopic campaigns. Indeed, with additional spectroscopic epochs, it becomes more likely to detect RV variability and identify binary systems, so our observed binary fraction is a lower limit.}

However, it should be noted that the majority of these systems are also pulsating stars and their significant RV variability could be caused by pulsations. \revision{Examples include TIC~53968952 and TIC~288489491, whose light curves and LS periodograms are shown in Fig.~\ref{fig:pulsators}, which have $\Delta$RV values of 44.5 and 20.5~km\,s$^{-1}$, respectively. Without photometric data, such systems would likely be labelled as binaries by default since their RV variability passed both binary detection criteria. Yet, their photometric variability is also quite large (43 and 93 mmag, respectively) and as such would be sufficient to cause large RV variability.}

\revision{After isolating} the list of pulsators identified using TESS light curves, the RV distribution for the sub-sample of non-pulsating binary candidates is shown in the bottom panel of Fig.~\ref{fig:RVdist}. For comparison, the distribution for the pulsating stars with significant RV variability is shown in the top panel. Therefore, the number of non-pulsating binaries that satisfy both criteria is 1~per~cent of the full sample. However, we emphasise that this is likely an underestimate of the true binary fraction because of different observational and astrophysical biases, which are discussed below.

First, \revision{while} we have excluded pulsators from the distribution in the bottom panel of Fig.~\ref{fig:RVdist}, it is important to note that galactic B-type stars are commonly both pulsators and found in binary systems \citep{burssens2020, southworth2022, eze2024}. However, the limited number of spectroscopic epochs for our sample means it is not possible to confidently identify if the significant RV variability arises from pulsations, binarity, or both. In our investigation, we have chosen to be conservative and assume that if a star is found to be a pulsator based on its TESS light curve then the spectroscopic RV variability could be caused by pulsations, which means it may not be a binary. Of course, this may be incorrect, since most stars may in fact be pulsating binaries and spectroscopy alone is difficult to establish this. In the ideal scenario, one should check if the candidate binary period found in spectroscopy matches a pulsation period identified in photometry taking all available information into account, such as range of expected pulsation periods given the star's location in the HR~diagram. For example, if a star has a significant spectroscopic binary period larger than about 20~d, then this is too large to be a heat-driven pulsation period in a main-sequence $\beta$~Cep star. On the other hand, if the identified spectroscopic period is of order a few days and similar periods are found in photometry indicating pulsations, this is a strong indication that the spectroscopic period is a pulsation period for such a star. Indeed, spectroscopic identification of pulsation periods, especially in B-type main-sequence stars, is a powerful technique for pulsation mode identification (see \citealt{aerts2010}).

Second, with only a few spectroscopic epochs, we are not sensitive to all possible binary orbital configurations. The selection function depends on the cadence and total time span of the observations, which is star dependent in our case and far from uniform across the sample. With an average cadence of the order of a month, and only two or three epochs, it is entirely possible to have missed a large range of orbital periods. Regardless, all systems with significant RV variability identified in this work are worthy of continued monitoring with spectroscopy to ascertain their binary status and the relationship to pulsations. Indeed, the interaction of binarity and pulsations --- tidal asteroseismology --- is an emerging field and very powerful for probing the physics of binary star evolution (see \citealt{southworth2025}).

\subsubsection{Choice of the 20~km\,s$^{-1}$ threshold}
\label{sec:threshold}

The $20$~km\,s$^{-1}$ threshold used in the second binary criteria (c.f. Eqn.~\ref{eq:rv2}) appears to align well with the overall $\Delta$RV distribution for pulsating and non-pulsating stars \revision{shown in the top and bottom panels of} Fig.~\ref{fig:RVdist}, \revision{respectively,} since we find the majority (79~per cent) of pulsators to be below this threshold. However, upon closer inspection, the $\Delta$RV distribution of pulsators (top panel of Fig.~\ref{fig:RVdist}) consists of three peaks and a long tail to large values. For example, a $20$~km\,s$^{-1}$ threshold fails to capture all of stars in the second peak ($10 \lesssim \Delta\mbox{RV} \lesssim 25$~km\,s$^{-1}$) nor any of the stars in the third peak ($25 \lesssim \Delta\mbox{RV} \lesssim 40$~km\,s$^{-1}$) in the distribution. We observe a non-negligible fraction of pulsators with $\Delta \mbox{RV}$ > 20~km\,s$^{-1}$ and an extended tail that reaches as high as 100~km\,s$^{-1}$. On the other hand, when considering only the non-pulsating stars (in the bottom panel of Fig.~\ref{fig:RVdist}), we find a far fewer systems with $\Delta\mbox{RV} > 20$~km\,s$^{-1}$ in absolute numbers but also a smaller relative fraction of the whole sample. 

\begin{figure}
    \centering
    \includegraphics[width=0.95\linewidth]{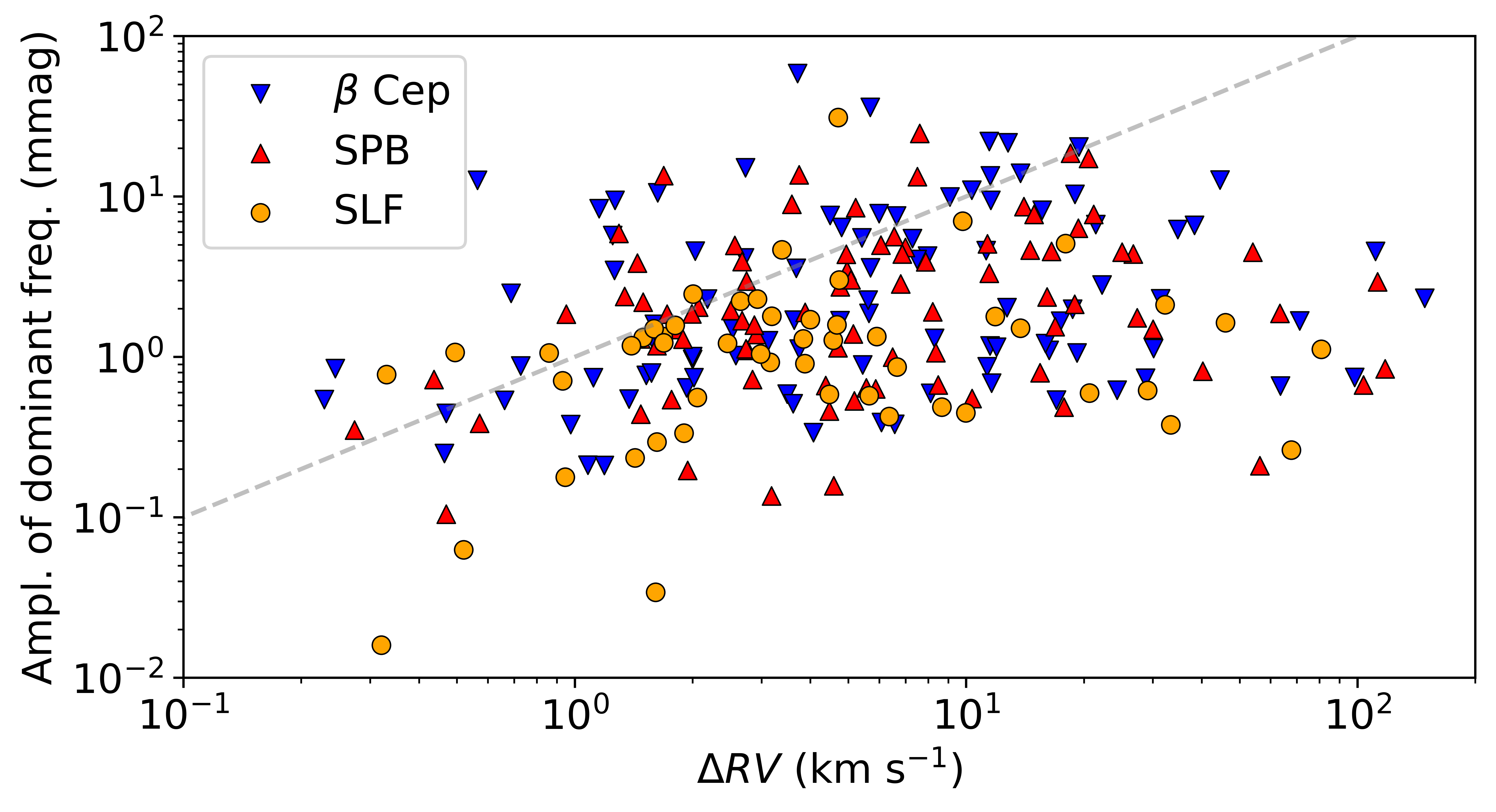}
    \caption{Comparison between calculated $\Delta \mbox{RV}$ values measured from multi-epoch HERMES spectra versus the amplitude of the dominant pulsation frequency based on TESS data, which are divided into the three main pulsator types. The dashed grey line is simply a unity line (i.e. not a fit) between the axis quantities (1 km\,s$^{-1}$ = 1~mmag).}
    \label{fig:RVvsAmp}
\end{figure}

To demonstrate the efficacy and validity of the $C=20$~km\,s$^{-1}$ threshold with respect to how pulsations among binary candidates can create false positives, we compare in Fig.~\ref{fig:RVvsAmp} the calculated $\Delta$RV values of each star to the highest-amplitude pulsation as measured in time-series TESS photometry. For the latter, we only considered frequencies larger than 1~d$^{-1}$ in an attempt to avoid including any periodicity potentially related to rotation. For visibility, we have added a dashed-grey unity line to Fig.~\ref{fig:RVvsAmp}, which represents that a spectroscopic pulsation amplitude of $\Delta$RV = 1~km\,s$^{-1}$ would be equivalent to a photometric amplitude of 1~mmag. Globally, we observe that the different pulsator types approximately follow the same trend, albeit with a lot of scatter, and no single pulsator type is over-represented in any one region of the parameter space. Calculating the Pearson correlation coefficients for each pulsator class, we get values of $-0.039$, $-0.038$, and $-0.042$, for the populations of $\beta$~Cep, SPB, and SLF respectively. The considerable scatter in Fig.~\ref{fig:RVvsAmp} demonstrates that a single threshold of $C=20$~km\,s$^{-1}$ for galactic B-type stars applied without any knowledge of the pulsation RV distribution may lead to mis-classification of high-amplitude pulsators as binaries and vice versa, in alignment with the findings of \citet{simondiaz2024}. 

On balance, the choice of 20~km\,s$^{-1}$ as a threshold seems reasonable for a large sample of galactic massive stars that contains a large fraction of pulsators. But there is a need for considerable caution for applying such a criterion blindly with no knowledge of the underlying pulsation fraction and as a consequence the pulsation-induced RV distribution. With additional spectroscopic epochs, we would expect the median $\Delta$RV value to increase as both additional pulsators and binaries are identified. On the other hand, shifting the $\Delta$RV threshold to larger values would decrease the number of false binary positives, but would also increase the number of false binary negatives. This is because genuine binaries are more likely to not be classified as a binary if using a higher RV threshold. Therefore, we conclude that an informed investigation of photometric pulsation periods in parallel to spectroscopic binary periods is highly advisable for studies of galactic massive stars (see also Nardini et al. in prep). In this way, an optimum choice of threshold can be made since it would be informed by photometric constraints on the amplitudes and periods of pulsations. Additionally, when considering the results of the various studies investigating binary fractions for a range of sample sizes and types found in the literature, a single threshold value seems unable to take into account the full breadth of variability observed within the populations of O- and B-type stars and that, following the conclusions of \citet{simondiaz2024}, one should rather consider multiple thresholds based on a given star's position in the HR diagram.

\section{Rotational modulation}
\label{sec:magnetism}

Rotational modulation is the periodic variation in stellar brightness caused by non-uniform surface features, such as chemical spots resulting from the presence of a large-scale magnetic field (e.g. \citealt{stibbs1950}). Since the chemical spots of early-type stars are long-lived, the photometric signature of rotational modulation is a series of harmonics in the LS periodogram with the base frequency being the rotation frequency (e.g. \citealt{bowman2018b, mathys2020}). Previous studies have yielded success rates between 50 and 70~per~cent for the detection of magnetic fields via spectropolarimetry for stars with chemical spots inferred from rotational modulation \citep[e.g.][]{buysschaert2018}. Rotational modulation also has the added benefit of being applicable to a wide range of spectral types, as compared to chemical peculiarity, whose features and causes can vary between spectral types (see \citealt{preston1974}). 

\begin{figure}
    \centering
    \includegraphics[width=0.95\linewidth]{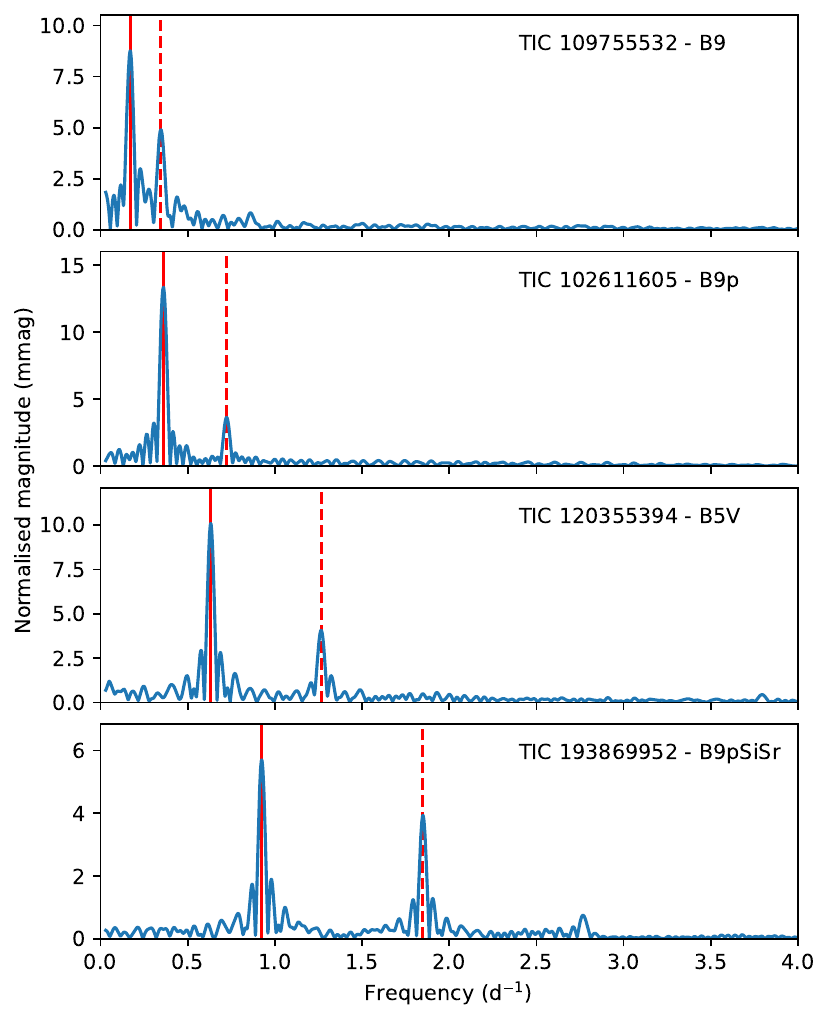}
    \caption{LS~periodograms for four example stars showing rotational modulation. In each case, the solid red line corresponds to the rotation frequency, and the dashed red line corresponds to its first harmonic ($f=2\times f_{\rm rot}$). The TIC~ID of each star and its SIMBAD spectral type is shown in the top-right corner of each panel.}
    \label{fig:rotmod}
\end{figure}

Based on all available TESS light curves for our sample and the associated frequency spectra, we searched for rotational modulation. We define a star to have rotational modulation if at least one harmonic of an isolated low-frequency signal is present in the frequency spectrum (i.e. two harmonically-related frequencies). A few examples of such stars with rotational modulation are shown in Fig.~\ref{fig:rotmod}. Using this criterion, we identify 148 stars in our sample with rotational modulation. This subset of stars may show multiple forms of variability, such as pulsating stars that also present rotational modulation, and is not necessarily these star's dominant form of variability. 
%In addition, this population includes an overlap of 75 stars which present both rotational modulation and CP properties via the 5200 \AA{} depression.
Therefore, these 148 stars with rotational modulation are worthy of follow-up spectropolarimetry to search for the presence of a large-scale surface magnetic field.

%\subsection{Pulsating candidate magnetic stars}

Of particular interest are candidate magnetic stars with pulsations, since these are prime targets for magneto-asteroseismology (see \citealt{buysschaert2018b, lecoanet2022}). We identify 
%seven candidate-magnetic CP stars with SPB pulsations and one showing $\beta$~Cep pulsations, as well as 
15 SPB and 16 $\beta$~Cep stars showing rotational modulation. 
%In addition, there is significant overlap between these subsets, with several pulsating stars with both rotational modulation and the 5200 \AA{} depression, which is strong indication that these stars are very likely magnetic pulsators. 
Although a modest sample in terms of size, we shall perform an in-depth investigation of these stars in a future work.

\section{Conclusions}

We have combined multi-epoch high-resolution spectra from a dedicated large programme with the HERMES spectrograph mounted on the Mercator telescope with new 2-min cadence TESS photometry to investigate the ensemble properties of a large sample of 873 galactic O- and B-type stars. In particular, we focus on spectroscopic and photometric signatures of pulsations, binarity, and rotational modulation as a potential indicator for magnetism. This much larger sample of Northern massive stars complements the previous study by \citet{burssens2020}, which was focussed on the Southern hemisphere. 

We demonstrate that the vast majority of massive galactic stars are pulsators, with a diversity of pulsations allowing us to identify $\beta$~Cep stars, SPB stars, and SLF variability. In general, photometric variability amongst massive stars appears to be nearly ubiquitous, with only 58 of the 873 stars in our sample showing no significant variability at all (i.e. constant stars). These 58 stars are all B-type stars but have a range of luminosity classes. With longer TESS light curves and/or improved photometric precision, it is not impossible that many of these would be detected as photometrically variable as well. 
%This number would very likely decrease in the cases where only a single HERMES spectra or TESS sector is available, and longer period variability is made accessible. 
As part of our ensemble analysis, we have determined the empirical locations of the $\beta$~Cep, SPB and SLF variability in the HR~diagram as the parameter space in which the 33rd and 66th percentile of stars show a specific type of variability. The stellar evolution tracks and corresponding theoretical instability regions for p and g~modes used in this work originate from \citet{burssens2020}, but assume a single set of physical prescriptions. Hence, these theoretical predictions do not capture changes in the location of the TAMS because of different mixing processes, nor differences caused by metallicity, rotation, or magnetism. Yet, the location of pulsating stars agrees fairly well with expectations. Most importantly, we have used the empirical instability regions of these three pulsator types to demonstrate that CBM is likely mass dependent \citep[e.g.][]{scott2021,whitehead2026}, especially above $M \gtrsim 15$~M$_{\odot}$.

Our multi-epoch HERMES spectra allowed us to identify 82 spectroscopic binaries, including 22 SB2 systems. We also identify the photometric signatures of binarity and discover 30 new eclipsing binaries, with 12 of them containing a $\beta$~Cep or SPB pulsator and 11 of them having SLF variability. 
%Pulsating EBs are ideal laboratories for improving stellar structure and evolution theory, since their model-independent dynamical masses and radii are powerful constraints in forward asteroseismic modelling -- see review by \citet{southworth2025}. 
We investigated the impact of the choice of RV threshold for identifying spectroscopic binary systems. We find that a non-negligible fraction of single pulsating stars have pulsation amplitudes larger than the typically used threshold of 20~km\,s$^{-1}$ for determining significant RV variability presumed to be caused by binarity. Whilst O-type stars in low-metallicity environments, such as the LMC and SMC galaxies, are typically not high-amplitude pulsators, this is not the case of galactic B-type stars. The distribution of pulsation amplitudes for galactic (single) B-type stars peaks at around 20~km\,s$^{-1}$ and has an extended tail that reaches upwards of 40~km\,s$^{-1}$. In our sample, we have the advantage of being able to directly compare the spectroscopic RV amplitudes and photometric amplitudes, thus providing additional insight into being able to distinguish between one source of RV variability from another 
%identify cases of false positive and false negative binaries 
when using a threshold of 20~km\,s$^{-1}$. Therefore, given the availability of TESS data across the entire sky, we advocate that spectroscopic studies of binarity among galactic massive stars should be informed by photometry. Unfortunately, the number of spectroscopic epochs for most stars in our sample is limited, and so a full binary analysis is not possible, but we have identified systems worthy of follow-up.

In addition, we identify 148 stars with rotational modulation in their TESS light curves, which is a proven indicator of a potential large-scale strong magnetic field for early-type stars \citep[e.g.][]{buysschaert2018,shultz2019}. 
%magnetic stars using two features that demonstrate strong reliability for identifying the presence of strong surface magnetism in the literature: rotational modulation in photometry and the 5200 \AA{} depression in spectroscopy. This is above the 10~per~cent incidence rate of strong ($>300$ G) surface magnetic fields typically reported for main-sequence OB(A)-type stars based on spectropolarimetry \citep{grunhut2017,shultz2019,sikora2019}. However, we reiterate that stars that show rotational modulation and/or the 5200~\AA{} depression are only magnetic candidates, and more in-depth follow-up and magnetic characterisation is required to properly assess this aspect. 
We also identify 39 pulsating stars with rotational modulation, which may have strong, large-scale magnetic fields making them potential targets for magneto-asteroseismology. The inclusion of magnetic fields in a forward asteroseismic modelling framework yields more robust insight into the rotation and mixing properties of magnetic stars.

% Finally, we find 127 stars showing emission features typical of OBe stars, of which 88 also show pulsation modes. Additionally, 23 of them show evidence of binarity (eclipsing or spectroscopic). 

% The high percentage of pulsators and low percentage of binarity for OBe stars is in agreement with previous work (e.g. \citealt{bodensteiner2020b, labadie2022}).

In the future, we will perform forward asteroseismic modelling of individual targets that have been characterised as high-priority from this work to calibrate the physical prescriptions of mixing in stellar structure and evolution models. While individual TESS sectors are usually sufficient to distinguish pulsator types, they are typically not sufficient for performing precise mode identification (see \citealt{scott2026}). The ideal scenario is, therefore, to have multiple consecutive sectors of TESS data, which is only realised for a fraction of our sample. Massive stars were generally avoided by the Kepler mission, and the light curves assembled by the K2 and TESS missions are, on average, quite short for robust forward asteroseismic modelling. However, we expect long-duration, short-cadence, and high-precision photometric precision light curves from the ESA PLATO mission \citep{rauer2025}, which is scheduled to be launched in early-2027. Thanks to PLATO's observing strategy, a long temporal baseline of at least 2~yr will enable pulsation mode identification for many massive stars. Our current study demonstrates that the asteroseismic return of studying massive stars with TESS data is very high, hence an equally high return is expected for massive stars observed by the PLATO mission. Combining these TESS photometric data with other observing techniques, such as spectroscopy and spectropolarimetry, we will form a more complete picture of how massive stars evolve and interact with their companions.

\section*{Acknowledgements}

This research has made use of the SIMBAD database operated at CDS, Strasbourg (France), and of NASA's Astrophysics Data System (ADS). 

Based on observations made with the Mercator Telescope (\url{https://www.mercator.iac.es}), operated on the island of La Palma by the Flemish Community, at the Spanish Observatorio del Roque de los Muchachos of the Instituto de Astrofísica de Canarias, and obtained with the HERMES spectrograph, which is supported by the Research Foundation - Flanders (FWO), Belgium, the Research Council of KU Leuven, Belgium, the Fonds National de la Recherche Scientifique (F.R.S.-FNRS), Belgium, the Royal Observatory of Belgium, the Observatoire de Genève, Switzerland and the Thüringer Landessternwarte Tautenburg, Germany. The authors thank the Institute of Astronomy (KU Leuven) and the many observers which contributed to the data gathering.

This paper includes data assembled by the TESS mission, with funding for the TESS mission provided by NASA's Science Mission directorate. This work used data from the European Space Agency (ESA) mission Gaia (\url{https://www.cosmos.esa.int/gaia}), processed by the Gaia Data Processing and Analysis Consortium (DPAC; \url{https://www.cosmos.esa.int/web/gaia/dpac/consortium}). Funding for the DPAC of the Gaia mission is provided by national institutions, in particular the institutions participating in the Gaia Multilateral Agreement.

The authors gratefully acknowledge UK Research and Innovation (UKRI) in the form of a Frontier Research grant under the UK government's ERC Horizon Europe funding guarantee (SYMPHONY; PI Bowman; grant number: EP/Y031059/1), and a Royal Society University Research Fellowship (PI Bowman; grant number: URF{\textbackslash}R1{\textbackslash}231631). This project received support from the ``La Caixa'' Foundation (ID 100010434) under the fellowship code LCF/BQ/PI23/11970035. JB is supported by an NWO Veni fellowship (VI.Veni.242.199). GH thanks the Polish National Center for Science (NCN) for support through grant 2021/43/B/ST9/02972. N.S. and I.A. acknowledge support from the Israel Science Foundation (grant number 2752/19) and from the Pazy foundation (grant number 216312). I.A. acknowledges further support from the European Research Council (ERC) under the European Union’s Horizon 2020 research and innovation program (grant agreement number 852097). TS acknowledges support from the Israel Science Foundation (ISF) under grant number 0603225041 and from the European Research Council (ERC) under the European Union's Horizon 2020 research and innovation program (grant agreement 101164755/METAL). S.S-D. acknowledges support from the State Research Agency (AEI) of the Spanish Ministry of Science and Innovation (MICIN) and the European Regional Development Fund, FEDER under grants PRODUCTOS DE LA INTERACCION DE ESTRELLAS MASIVAS REVELADOS POR GRANDES SONDEOS ESPECTROSCOPICOS, with reference PID2024-159329NB-C21. S.S-D. also acknowledge funding from European Commission (EC) under Project OCEANS - Overcoming challenges in the evolution and nature of massive stars, HORIZON-MSCA-2023-SE-01, No G.A~101183150 Funded by the European Union. Views and opinions expressed are however those of the author(s) only and do not necessarily reflect those of the European Union or the European Research Executive Agency (REA). Neither the European Union nor the granting authority can be held responsible for them. \revision{AT acknowledges support from the BELgian federal Science Policy Office (BELSPO) through PRODEX grant PLATO (ZKE8588), from the Flemish Government under the long-term structural Methusalem funding program by means of the project SOUL: Stellar evolution in full glory, grant METH/24/012 at KU Leuven, and from the Research Foundation – Flanders (FWO) (grant agreement G0ABL24N).}

%%%%%%%%%%%%%%%%%%%%%%%%%%%%%%%%%%%%%%%%%%%%%%%%%%
\section*{Data Availability}
 
The TESS data used in this work are publicly available via the MAST website: \url{https://archive.stsci.edu/missions-and-data/tess}. \revision{The HERMES data used in this work are publicly available via the Mercator Observatory Archive website: \url{https://www.mercator.iac.es/instruments/hermes/archive/}}. The GAIA data used in this work are publicly available via the Gaia website: \url{https://gea.esac.esa.int/archive/}, processed by the Gaia Data Processing and Analysis Consortium (DPAC; \url{https://www.cosmos.esa.int/web/gaia/dpac/consortium}).

This research has made use of the following open-access software packages: {\tt TLUSTY} for the grids of synthetic spectra \citep{lanz2003,lanz2007}, {\tt LIGHTKURVE} (\url{https://lightkurve.github.io/lightkurve/}), a {\tt PYTHON} package for \emph{Kepler} and TESS data analysis (Lightkurve Collaboration 2018), {\tt PERIOD04} (\url{https://www.period04.net}) for frequency analysis \citep{lenz2005}, as well as {\tt matplotlib} \citep{hunter2007}, {\tt numpy} \citep{harris2020array}, {\tt astropy} \citep{astropy1, astropy2, astropy3}, and {\tt scipy} \citep{2020SciPy-NMeth}.

For the purpose of open access, the authors have applied a CC BY licence to the author accepted manuscript version: \url{https://arxiv.org/abs/TBD}. Data products that support the results in this paper are publicly available via the Zenodo repository: \url{https://zenodo.org/records/10.5281/zenodo.21414429}.



\bibliographystyle{mnras}
\bibliography{bibliography} % if your bibtex file is called example.bib

%%%%%%%%%%%%%%%%%%%%%%%%%%%%%%%%%%%%%%%%%%%%%%%%%%

%%%%%%%%%%%%%%%%% APPENDICES %%%%%%%%%%%%%%%%%%%%%

\appendix

\clearpage
% {\onecolumn
\section{Comparison of continuum normalisation technique}

\begin{figure*}
    \centering
    \includegraphics[width=0.95\linewidth]{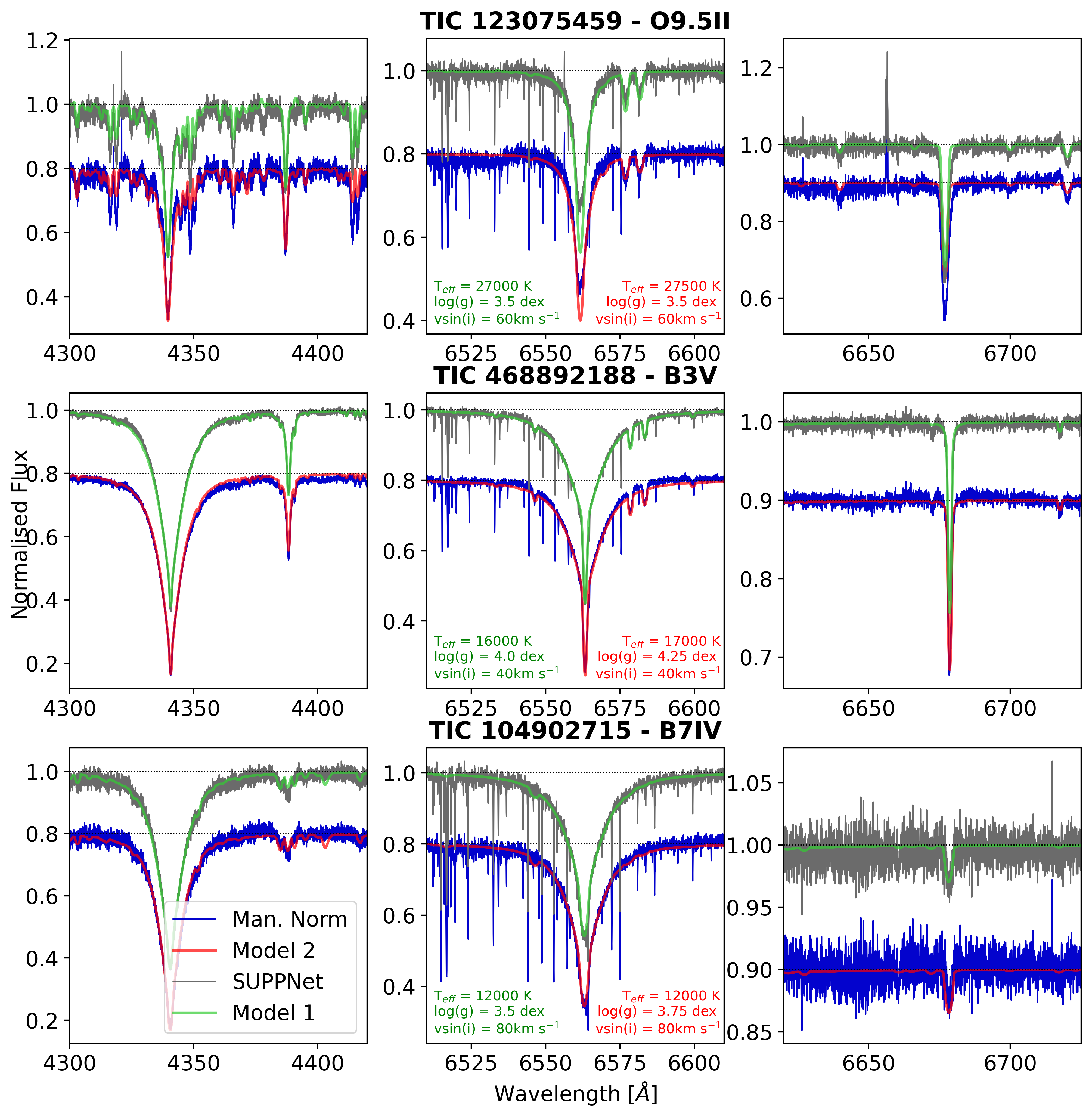}
    \caption{Comparison of the results of normalisation between manual (blue) and via \textsc{SUPPNet} (grey) for three representative stars. The resultant best-fitting TLUSTY models are shown in red and green, respectively, and the best-fitting parameters of each solution are shown in same colour in the middle panel for each row. A dashed line has been added to each spectrum to show the continuum level.}
    \label{fig:contnorm}
\end{figure*}

In Fig.~\ref{fig:contnorm}, we show a comparison of the best-fitting TLUSTY models derived from the manually normalised and {\sc SUPPNet}-normalised HERMES spectra, for a selection of stars. This comparison, which is typical for the entire sample, demonstrates the excellent result of {\sc SUPPNet}, since the difference in $T_{\rm eff}$ and $\log\,g$ values between each best-fitting model is usually a maximum of one grid step size.

\section{New eclipsing binary light curves}

\begin{figure*}
    \centering
    \includegraphics[width=0.95\linewidth]{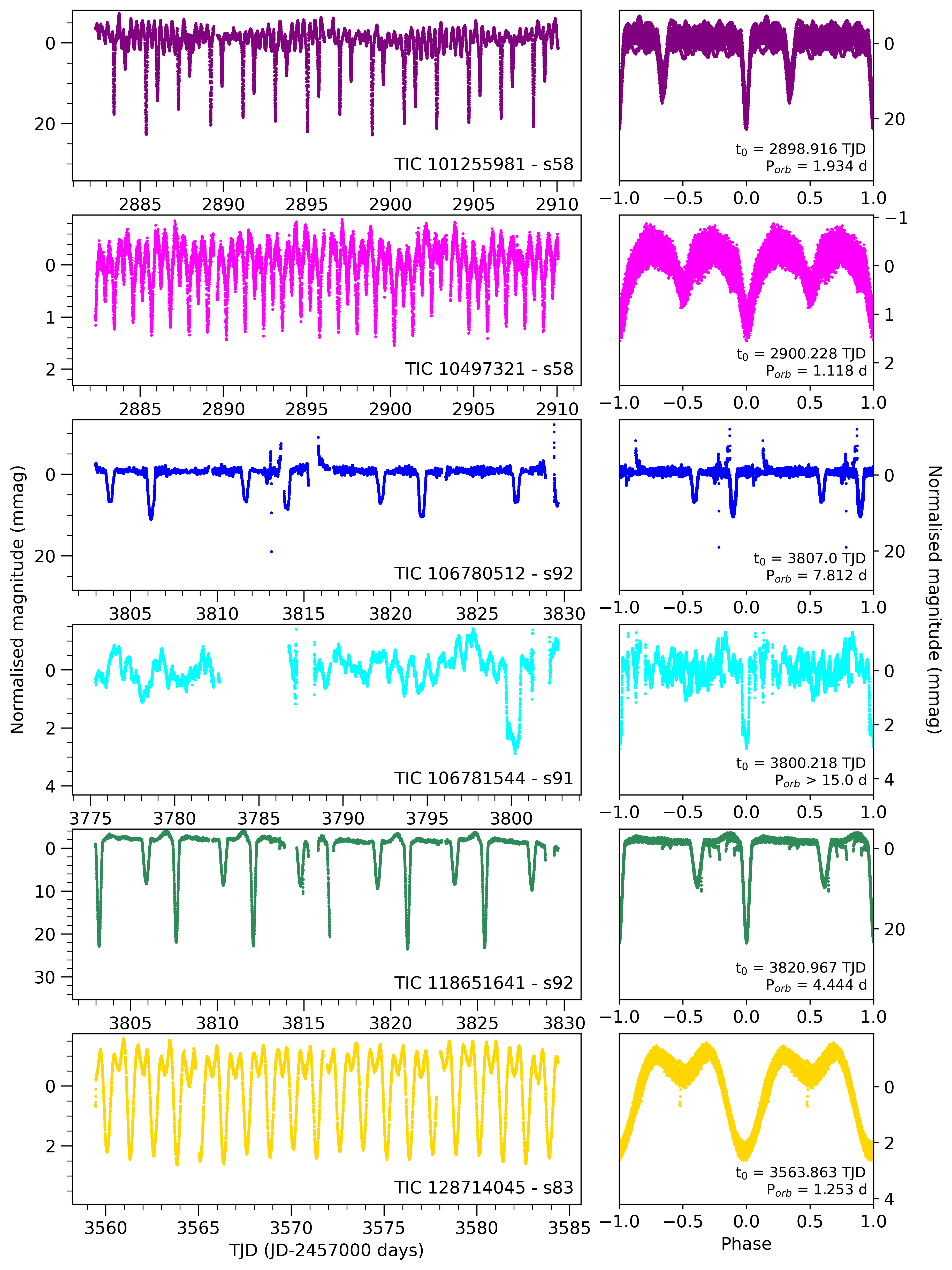}
    \caption{Left: Light curves for the 30 new EBs identified through this work, with their TIC ID and respective sector number. Right: Phase-folded light curves, with the corresponding orbital period P$_{\text{orb}}$ and reference time T$_0$ based on the primary eclipse.}
    \label{fig:EB_1}
\end{figure*}

\begin{figure*}
    \centering
    \includegraphics[width=0.95\linewidth]{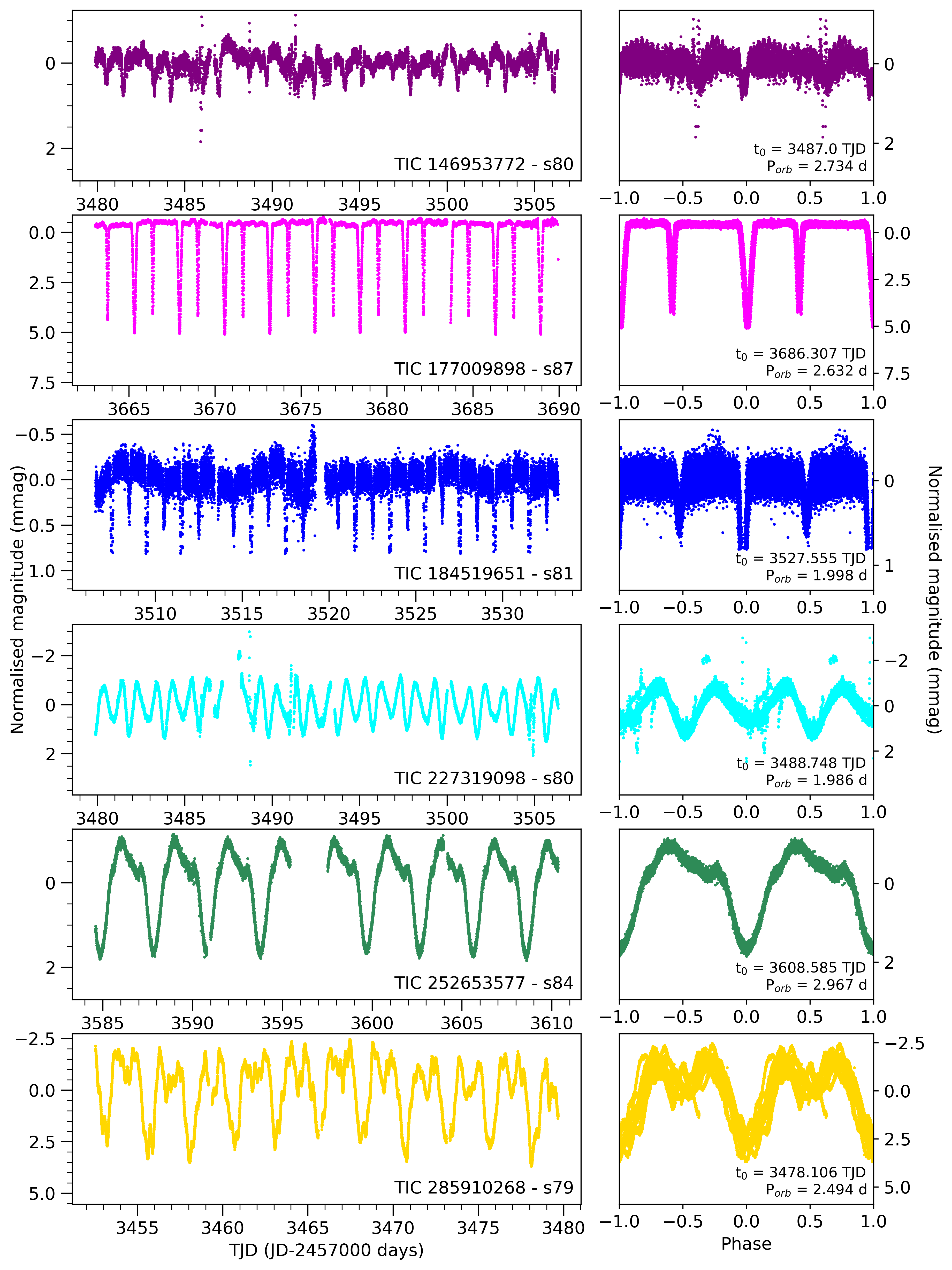}
    \caption{Same as Fig.~\ref{fig:EB_1} (continued.).}
    \label{fig:EB_2}
\end{figure*}

\begin{figure*}
    \centering
    \includegraphics[width=0.95\linewidth]{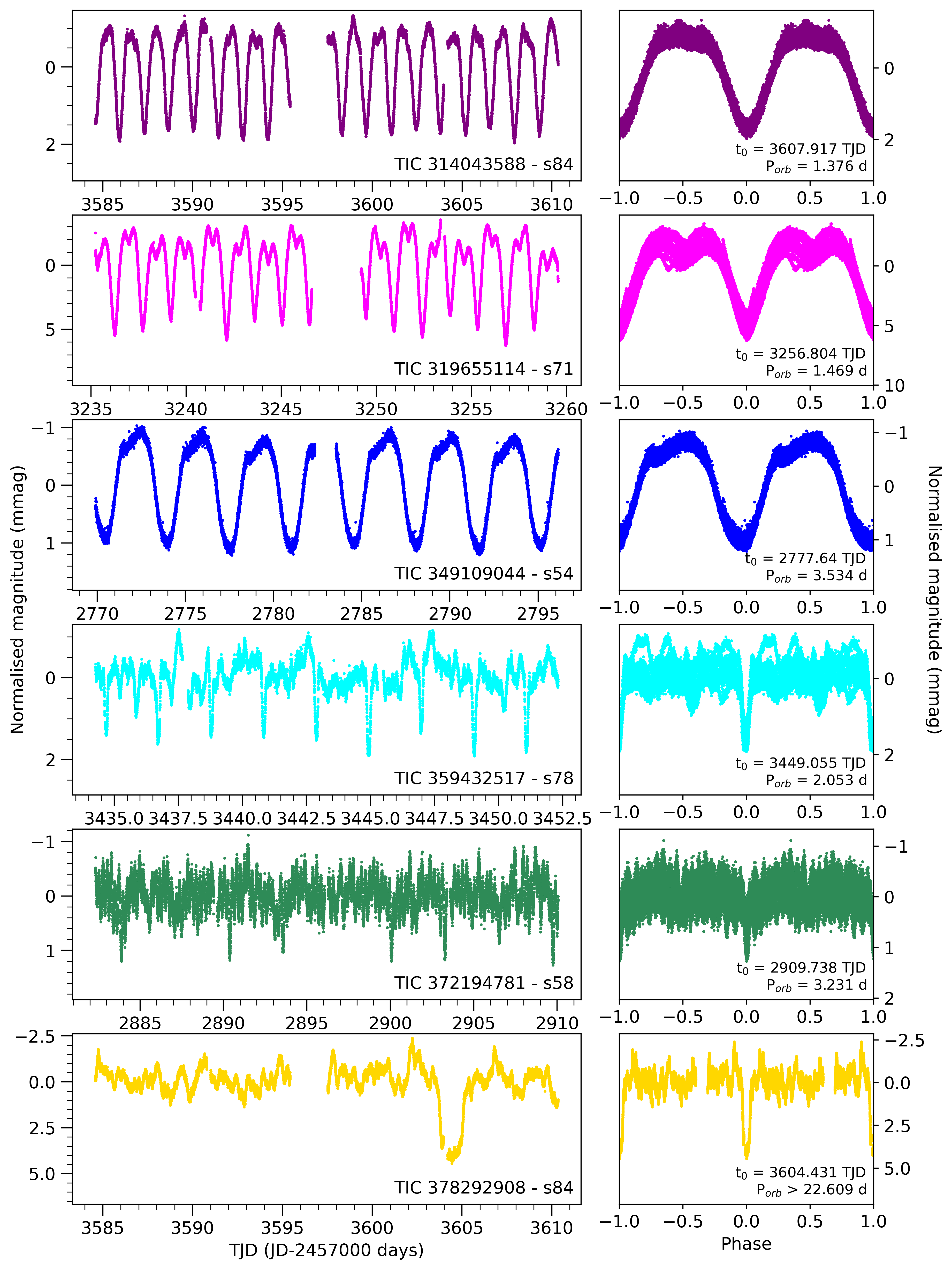}
    \caption{Same as Fig.~\ref{fig:EB_1} (continued.).}
    \label{fig:EB_3}
\end{figure*}

\begin{figure*}
    \centering
    \includegraphics[width=0.95\linewidth]{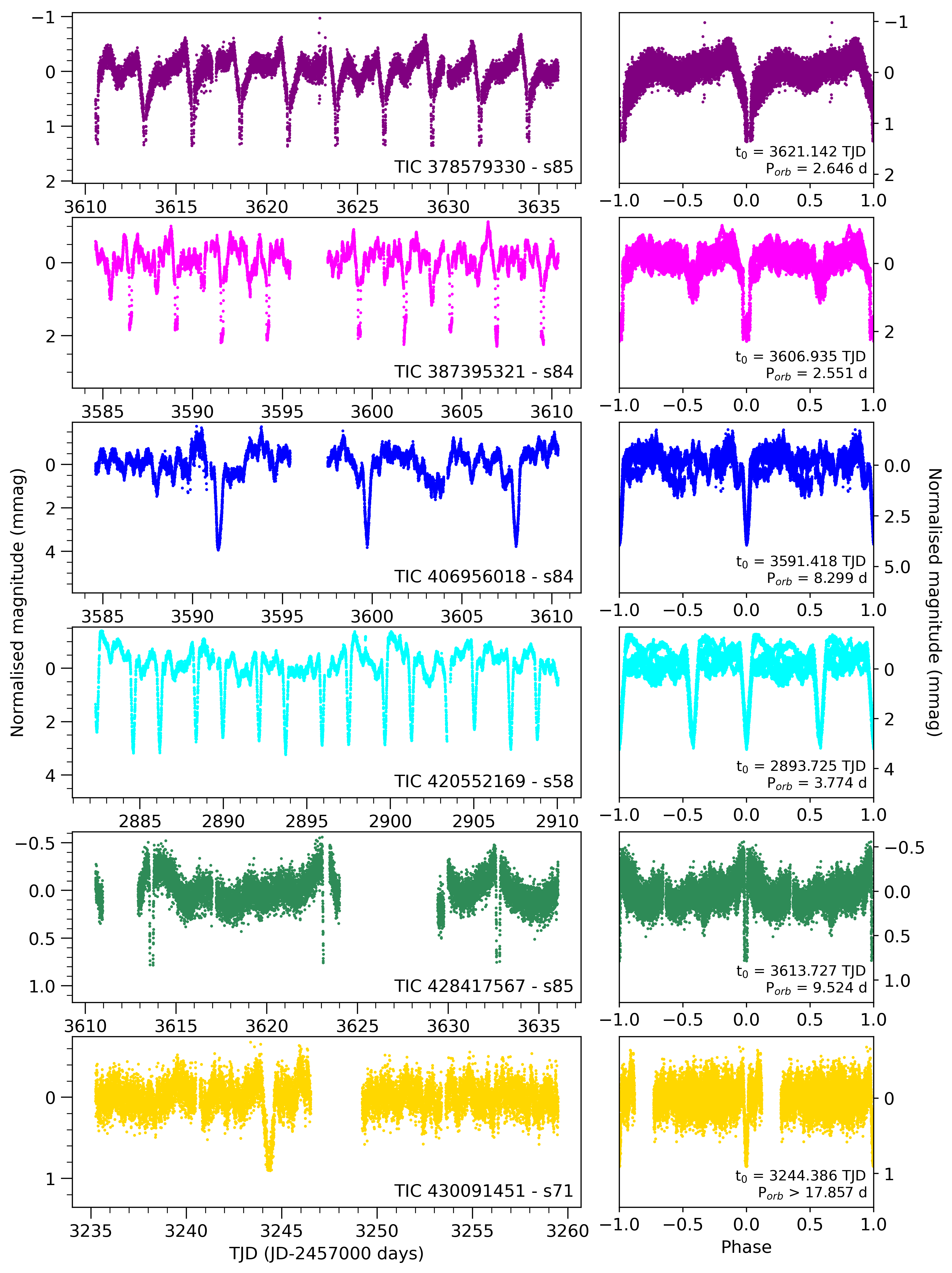}
    \caption{Same as Fig.~\ref{fig:EB_1} (continued.).}
    \label{fig:EB_4}
\end{figure*}

\begin{figure*}
    \centering
    \includegraphics[width=0.95\linewidth]{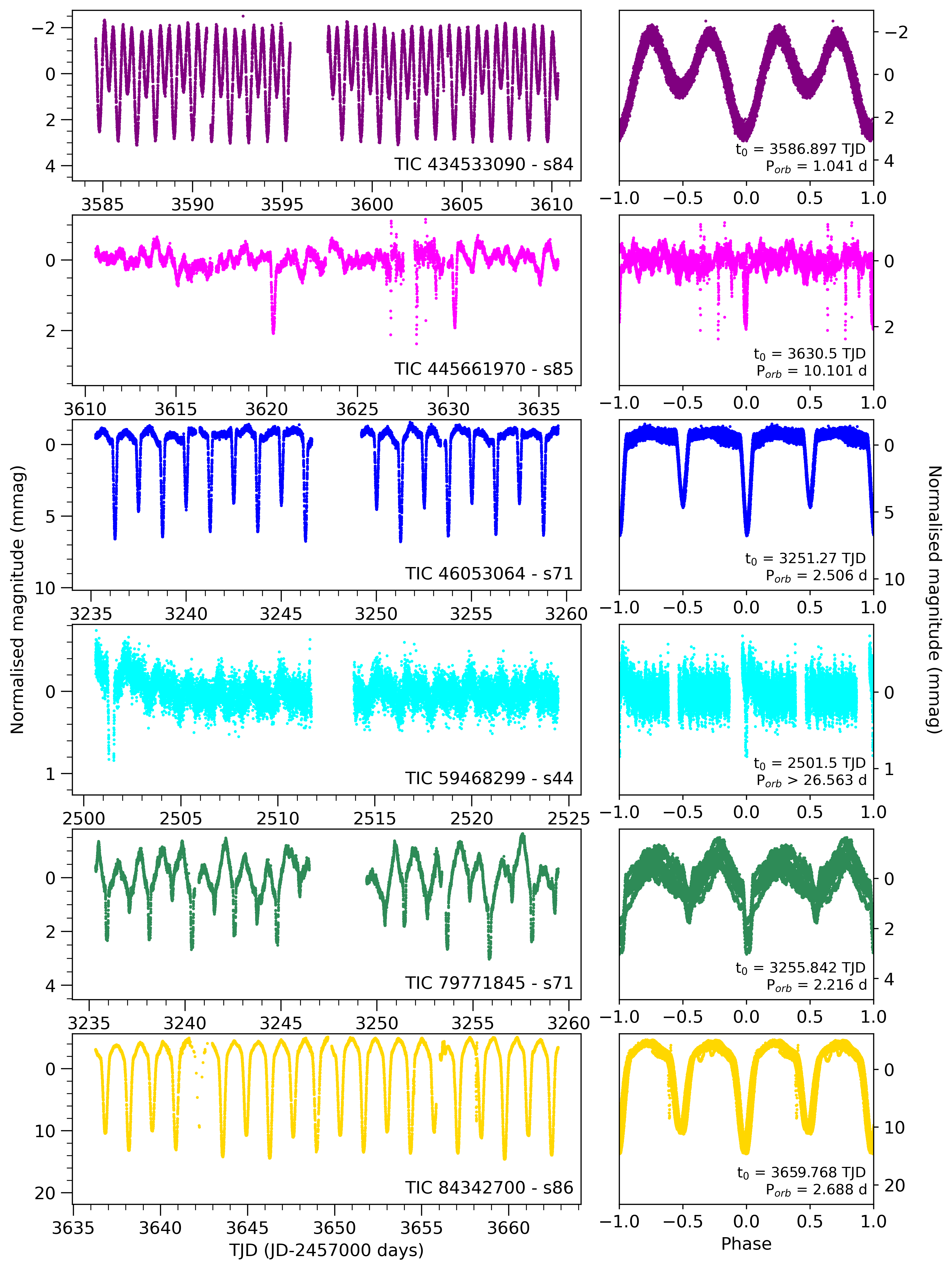}
    \caption{Same as Fig.~\ref{fig:EB_1} (continued.).}
    \label{fig:EB_5}
\end{figure*}

In Figs.~\ref{fig:EB_1}--\ref{fig:EB_5}, we show TESS light curves of all the new EBs that have been discovered in this work.

% }
%\section{Candidate magnetic pulsators}

%%%%%%%%%%%%%%%%%%%%%%%%%%%%%%%%%%%%%%%%%%%%%%%%%%

% Don't change these lines
\bsp	% typesetting comment
\label{lastpage}
\end{document}